\documentclass[final,3p,times,twocolumn,fleqn]{elsarticle}

\usepackage{epstopdf}
\usepackage{amsmath}

\usepackage[caption=false]{subfig}
\usepackage{amssymb}

\journal{Computers \& Fluids}

\begin{document}

\begin{frontmatter}

\title{Fully dissipative relativistic lattice Boltzmann method in two dimensions}


\author[label1,label2]{Rodrigo C. V. Coelho}
\ead{rcvcoelho@if.ufrj.br}

\author[label2]{Miller Mendoza}
\ead{mmendoza@ethz.ch}

\author[label1]{Mauro M. Doria}
\ead{mmd@if.ufrj.br}

\author[label2]{Hans J. Herrmann}
\ead{hans@ifb.baug.ethz.ch}

\address[label1]{Departamento de F\'{\i}sica dos S\'{o}lidos, Universidade Federal do Rio de Janeiro, 21941-972 Rio de Janeiro, Brazil}%
\address[label2]{ ETH Z\"{u}rich, Computational Physics for Engineering Materials, Institute for Building Materials, Schafmattstrasse 6, HIF, CH-8093 Z\"{u}rich, Switzerland}

\begin{abstract}
In this paper, we develop and characterize the fully dissipative Lattice Boltzmann method for ultra-relativistic fluids in two dimensions using three equilibrium distribution functions: Maxwell-J\"uttner, Fermi-Dirac and Bose-Einstein. Our results stem from the expansion of these distribution functions up to fifth order in relativistic polynomials. We also obtain new Gaussian quadratures for square lattices that preserve the spatial resolution. Our models are validated with the Riemann problem and the limitations of lower order expansions to calculate higher order moments are shown. The kinematic viscosity and the thermal conductivity are numerically obtained using the Taylor-Green vortex and the Fourier flow respectively and these transport coefficients are compared with the theoretical prediction from Grad's theory. In order to compare different expansion orders, we analyze the temperature and heat flux fields on the time evolution of a hot spot.
\end{abstract}


\end{frontmatter}



\section{Introduction}

Relativistic fluid dynamics and kinetic theory~\cite{cercignani02} for relativistic gases play an important role in the study of many physical systems, ranging from the big scale of cosmology and astrophysics~\cite{shore12,bernstein88,0143-0807-36-1-015007} to the microscopic scale of particle physics~\cite{kouno90} and condensed matter physics~\cite{PhysRevLett.103.025301}. Since the discovery of the elliptic flow in the quark-gluon plasma formed by ultra-relativistic nuclear collisions~\cite{Ackermann01}, several models based on relativistic fluid dynamics have appeared to describe the experimental results~\cite{hupp11,teaney09, bouras09}. Two-dimensional models have been used to describe galaxy formation~\cite{jones76}, cosmological models~\cite{kremer02} and quark-gluon plasma~\cite{SCHUKRAFT13,ollitralt92}, but they have gained special importance after the discovery of graphene~\cite{novoselov04, novoselov05} and Dirac materials~\cite{wehling14}, in which the charge carriers, governed by the Fermi-Dirac (FD) distribution, behave effectively as massless ultra-relativistic quasi-particles moving at Fermi speed. Relativistic models based on Bose-Einstein (BE) statistics have also applicability to explain fluid dynamics effects, as the collective behavior of the matter formed shortly after nuclear collisions~\cite{blaizot12}, gravitational analogy with BE condensates~\cite{fagnocchi10} and even attempts to explain dark matter~\cite{bohmer07}.

The lattice Boltzmann method~\cite{succi01} (LBM) is a numerical technique based on the Boltzmann equation and on the Gaussian quadrature, which have been successfully applied to model classical fluids~\cite{kruger16,coelho16}, governed by the Maxwell-Boltzmann (MB) distribution, and also to semi-classical~\cite{coelho14, coelho16-2, yang09} and relativistic fluids. For classical fluids, it has been demonstrated that the hydrodynamic equations can be fully recovered by the LBM if the equilibrium distribution function (EDF) is expanded in orthogonal polynomials up to a minimum order that retains the necessary moments~\cite{shan2006kinetic, 0295-5075-81-3-34005}. For instance, to recover the Navier-Stokes equation (momentum conservation equation) one needs to expand the MB distribution up to third order in Hermite polynomials and to recover the energy conservation equation the fourth order expansion is required (analogous results were found for semiclassical fluids~\cite{2017arXiv171009472C}). For relativistic fluids, one needs a fifth order expansion to fully recover the hydrodynamic equations and transport coefficients as shown in Refs.~\cite{mendoza13-2} and~\cite{mendoza13-3}.

In 2010, the first relativistic lattice Boltzmann method (RLBM) was proposed by Mendoza et. al.~\cite{mendoza10, mendoza10derivation} and subsequently improved in numerical stability and new features~\cite{hupp11, Succi2014relativistic, Debus2017energy, mohseni15}. The theoretical background for the RLBM and the extension to ultra-relativistic gases was done by Romatschke et. al.~\cite{romatschke11}, where the authors used a model with interpolated streaming since the velocity vectors disposed along a sphere do not match the square lattice. These first models were based on the expansion of the Maxwell-J\"uttner (MJ) distribution in relativistic polynomials up to second order following an analogous procedure as for classical models. In Ref.~\cite{mendoza13-3}, an improved dissipation model based on a third order expansion of the MJ distribution was proposed, which does not recover the dissipation completely because a fifth order expansion is required. This model relies on a new Gaussian quadrature with exact streaming on a square lattice, recovering one of the main advantages of LBM, but costing a loss of resolution. Recently, a new RLBM, also based on a third order expansion of the MJ distribution, was able to implement exact streaming on a square lattice without loosing spatial resolution allowing also to treat the regime of massive particles~\cite{gabbana17-2}. Meanwhile, other RLBMs with exact streaming have been used for graphene, where the grid points are disposed on a hexagonal lattice~\cite{oettinger13, furtmaier15} such as in the molecular structure of graphene. Nevertheless, for these quadratures, the polynomial expansion of the EDF is limited to second order, which might be enough for practical purposes, but gives a poor description if the velocities and/or the temperature fluctuations are moderately high, as shown here. In Ref.~\cite{coelho17}, the first model based on a fifth order expansion of the FD distribution was used to study the Kelvin-Helmholtz instability on graphene. Since this is a viscous fluid dynamical effect, a fully dissipative method is desirable to achieve better accuracy of the results. 

The Grad~\cite{grad49} and Chapman-Enskog~\cite{chapman70} (CE) methods are the most common ones to calculate, from the Boltzmann equation, the transport coefficients and the hydrodynamic equations, i.e, the Navier-Stokes equation and the Fourier law. Both methods give the same results in the non-relativistic case~\cite{kremer10}. For relativistic fluids, these two methods give the same conservation equations, but different transport coefficients~\cite{cercignani02}, which is still a controversial topic nowadays~\cite{bhalerao14}. Other methods have been proposed to obtain the transport coefficients in relativistic fluids, as the renormalization group method, which gives the same results obtained with the CE method~\cite{tsumura15, tsumura12, kikuchi16, PhysRevD.85.114047}. Numerical methods based on a bottom-up construction are important tools to gain insight about the correct form of the transport coefficients and recent simulations with three dimensional relativistic methods have consistently confirmed the prediction of the CE method~\cite{gabbana17, florkowski13, bhalerao14}. A careful analysis of wave attenuation in a medium formed by ultra-relativistic particles can be found in Ref.~\cite{PhysRevC.97.024914}, which exactly recovers the transport coefficients predicted by the CE method. Surprisingly, few have calculated the transport coefficients in two dimensions, despite the increasing importance that the two-dimensional relativistic systems have gained during the last years. In Ref.~\cite{mendoza13-2} the Grad method was applied in two dimensions,
obtaining the correspondent bulk and shear viscosities and the thermal conductivity, which have been used to calculate the relaxation time in RLBM simulations~\cite{oettinger13, mendoza13-3,mohseni15}. However, in Ref.~\cite{furtmaier15}, the shear viscosity and the thermal conductivity, numerically measured using a RLBM for graphene, disagree with the ones obtained with Grad's expansion. Furthermore, the bulk viscosity was also calculated in Ref.~\cite{kremer02} for a two-dimensional relativistic gas, but they did not investigate the other transport coefficients. So far, the CE method for relativistic gases in two dimensions has never been used to derive the full set of transport coefficients, remaining this an important open task to achieve a better understanding of the two-dimensional relativistic gases. 

In this paper, we build two-dimensional RLBMs with full dissipation using the MJ and BE distributions and compare them with the model for the FD distribution described in Ref.~\cite{coelho17}. To do so, we expand the EDFs up to fifth order and develop new Gaussian quadratures able to calculate tensors up to fifth order and that preserve the spatial resolution. To test and characterize our models, we perform five numerical simulations. The models are validated using the Riemann problem though the comparison of our results with the ones from a reference model (Ref.~\cite{mendoza13-3}). At this point, we calculate tensors from second to fifth order and show that different expansion orders give different results. We numerically measure, with high precision, the kinematic viscosity, with the Taylor-Green vortex, and the thermal conductivity, through Fourier flow and compare the results with the reference model. As will be shown, these measurements do not agree with the coefficients obtained with Grad's expansion, but they agree with previous measurements using RLBM~\cite{coelho17,furtmaier15}. The accuracy of our models regarding space discretization is verified in the context of forced Poiseuille flow. In addition, we simulate the hot-spot relaxation in order to observe the differences between different expansion orders for the temperature and heat flux fields. 

This work is organized as follows. In Secs. \ref{boltzmann-sec} and \ref{polynomials-sec} we review the relativistic Boltzmann equation and the relativistic orthogonal polynomials. In Sec. \ref{expansion-sec}, we describe the expansion of the three distribution functions (MJ, FD and BE) in orthogonal polynomials, with more details given in the Appendix, and in Sec. \ref{quadrature-sec} we calculate the Gaussian quadratures. The full EDF expansion of the distribution functions as well as the quadratures with high precision can be found in the Supplemental Material\footnote{See Supplemental Material at [URL will be inserted by publisher] for more details about the models.}. In Sec. \ref{numerical-sec} we describe and show the results for the five numerical tests used to validate and characterize the models. In Sec. \ref{conclusion-sec} we summarize the main results and conclude.

\section{Model description}

\subsection{Relativistic Boltzmann Equation}\label{boltzmann-sec}

The temporal evolution in our model is given by the relativistic Boltzmann equation~\cite{cercignani02} with the Anderson-Witting collision operator, which assumes a single relaxation time for the problem, $\tau$, and allows us to treat massless ultra-relativistic particles:
\begin{flalign}\label{boltzmann-eq-general}
\bar p^\mu\partial_\mu f =   - \frac{\bar p_\mu U^\mu}{c^2\tau}(f - f^{eq}),
\end{flalign}
where $c$ is the speed of light. Repeated indexes represent a sum (Einstein's notation). For two-dimensional systems, the greek indexes range from 0 to 2 (0 is the temporal component) while the latin indexes range from 1 to 2. The relativistic (2+1)-momentum stands for $\bar p^\mu = (E/c, \mathbf{\bar p})$, the (2+1)-velocity for $U^\mu = \gamma(u) (c, \mathbf{u})$ and the space-time coordinates for $x^\mu=(ct, \mathbf{x})$, where $\gamma (u) = 1/\sqrt{1-u^2/c^2}$ is the Lorentz factor. Note that Eq.~\eqref{boltzmann-eq-general} becomes the non-relativistic Boltzmann equation in the classical limit, $u/c \ll 1$. The EDF, $f^{eq}$, can be either Fermi-Dirac, Maxwell-J\"uttiner or Bose-Einstein distributions, as described in section \ref{expansion-sec}. 

In the ultra-relativistic limit, i.e., when the kinetic energy is much larger than the rest mass energy, $\bar p^\mu \bar p_\mu =  (\bar p^0)^2 - \mathbf{\bar p}^2 =0 \:\:\Rightarrow\:\: \bar p^0 = |\mathbf{\bar p}| $, and Eq. \eqref{boltzmann-eq-general} becomes
\begin{flalign}\label{boltz-eq-ultra-rel-eq}
 \frac{\partial f}{\partial t} + \mathbf{\mathbf{v}}\cdot \nabla f = - \gamma(1  -\mathbf{ \mathbf{v}}\cdot \mathbf{u}) \frac{(f-f^{eq})}{\tau },
 \end{flalign}
where $\mathbf{v} = \mathbf{\hat{ p}} =  \mathbf{\bar p}/|\mathbf{\bar p}| $ is the microscopic velocity with norm $c$, and we adopt from now on natural units: $c=k_B=\hbar=1$. In the numerical algorithm, the discrete form of Eq.\eqref{boltz-eq-ultra-rel-eq} is used:
\begin{flalign}
&f_\alpha(t+\delta t, \mathbf{r}+ \mathbf{\mathbf{v}}_\alpha \delta t) - f_\alpha(t,\mathbf{r}) \\ \nonumber
&= -\gamma (1-\mathbf{\mathbf{v}}_\alpha\cdot \mathbf{u})\frac{\delta t (f_\alpha - f_{\alpha}^{eq})}{\tau},
\end{flalign}
where $\delta t$ is the time step of the simulations. Because all particles move at (or nearly) the speed of light, the microscopic velocity is always unitary in natural units, $\vert \mathbf{v}_\alpha \vert = \vert \mathbf{\bar p}_\alpha/|\mathbf{\bar p}_\alpha| \vert= 1$, what must be considered in the Gaussian quadrature calculation, as done in section \ref{quadrature-sec}.  

\subsection{Relativistic Polynomials}\label{polynomials-sec}

In the LBM, the EDF is expanded in orthogonal polynomials up to a finite order $N$ in order to use the Gaussian quadrature~\cite{abramowitz64} for the exact equivalence between the sums and integrals. In this procedure, the information contained in the terms of order above $N$ is lost. Thus, to have a faster convergence in the expansion and, therefore, minimize the loss of accuracy due to the truncation, the weight function used in the orthogonalization of the polynomials should be close to the EDF. With this purpose, we calculate relativistic generalized polynomials following the procedure developed in Ref.~\cite{coelho16-2} for non-relativistic polynomials. The polynomials below allow us to find orthogonal polynomials for generic weight functions, $\omega(p)$.
\begin{flalign*}
 P = A_1, \:\:\:\:  P^{i_1} = B_1 p^{i_1}, \:\:\:\: P^{0} = C_1 p^0 + C_2
\end{flalign*}
 \begin{flalign*}
 P^{i_1i_2} =  D_1 p^{i_1}p^{i_2} + [D_2 (p^0)^2 + D_3 p^0 + D_4]\delta^{i_1i_2}, 
\end{flalign*}
 \begin{flalign*}
P^{i_10} =  [E_1 p^0 + E_2] p^{i_1},
\end{flalign*}
\begin{flalign*}
&P^{i_1i_2i_3} = F_1 p^{i_1}p^{i_2}p^{i_3} + [F_2 (p^0)^2 + F_3 p^0 + F_4]\\  &\cdot[p^{i_1}\delta^{i_2i_3} + p^{i_2}\delta^{i_1i_3} + p^{i_3}\delta^{i_1i_2}] ,
 \end{flalign*}
\begin{flalign*}
 & P^{i_1i_20} =  [G_1 p^0 + G_2] p^{i_1}p^{i_2}  + \delta^{i_1i_2}[G_3 (p^0)^3 \\ & + G_4 (p^0)^2 + G_5 p^0 + G_6]
 \end{flalign*}
\begin{flalign*}
 &P^{i_1i_2i_3i_4} =  H_1 p^{i_1}p^{i_2}p^{i_3}p^{i_4} + [H_2 (p^0)^2 +H_3 p^0 + H_4]\\ & \cdot [p^{i_1}p^{i_2}\delta^{i_3i_4}
 +p^{i_1}p^{i_3}\delta^{i_2i_4} +p^{i_1}p^{i_4}\delta^{i_2i_3}+ p^{i_2}p^{i_3}\delta^{i_1i_4}\\ &+ p^{i_2}p^{i_4}\delta^{i_1i_3}+ p^{i_3}p^{i_4}\delta^{i_1i_2}] + [H_5 (p^0)^4 + H_6 (p^0)^3 \\ & + H_7 (p^0)^2 + H_8 p^0 + H_9 ]\delta^{i_1i_2i_3i_4}
\end{flalign*}
\begin{flalign*}
 &P^{i_1i_2i_30} = [I_1 p^0 + I_2]p^{i_1}p^{i_2}p^{i_3} + [I_3 (p^0)^3 + I_4 (p^0)^2 \\ & + I_5 p^0 + I_6]
[p^{i_1}\delta^{i_2i_3}+p^{i_2}\delta^{i_1i_3}+p^{i_3}\delta^{i_1i_2}] 
 \end{flalign*}
\begin{flalign*}
 &P^{i_1i_2i_3i_4i_5} =  J_1 p^{i_1}p^{i_2}p^{i_3}p^{i_4}p^{i_5} + [J_2 (p^0)^2+J_3 p^0 +J_4]\\ & \cdot [p^{i_3}p^{i_4}p^{i_5}\delta^{i_1i_2} + p^{i_2}p^{i_4}p^{i_5}\delta^{i_1i_3} +p^{i_2}p^{i_3}p^{i_5}\delta^{i_1i_4} \\ & +p^{i_2}p^{i_3}p^{i_4}\delta^{i_1i_5} +p^{i_1}p^{i_4}p^{i_5}\delta^{i_2i_3}
+ p^{i_1}p^{i_3}p^{i_5}\delta^{i_2i_4}\\ & +p^{i_1}p^{i_3}p^{i_4}\delta^{i_2i_5} +p^{i_1}p^{i_2}p^{i_5}\delta^{i_3i_4}+p^{i_1}p^{i_2}p^{i_4}\delta^{i_3i_5} \\ & +p^{i_1}p^{i_2}p^{i_3}\delta^{i_4i_5}]+[J_5 (p^0)^4 + J_6 (p^0)^3 + J_7 (p^0)^2 \\ & + J_8 p  + J_9][p^{i_1}\delta^{i_2i_3i_4i_5} +p^{i_2}\delta^{i_1i_3i_4i_5}+p^{i_3}\delta^{i_1i_2i_4i_5} \\ & +p^{i_4}\delta^{i_1i_2i_3i_5}+p^{i_5}\delta^{i_1i_2i_3i_4}] 
\end{flalign*}
\begin{flalign*}
 &P^{i_1i_2i_3i_40} =  [K_1 p^0 + K_2]p^{i_1}p^{i_2}p^{i_3}p^{i_4} + [K_3 (p^0)^3 \\ &+K_4 (p^0)^2 + K_5 p^0 + K_6]
[p^{i_3}p^{i_4}\delta^{i_1i_2}  + p^{i_2}p^{i_4}\delta^{i_1i_3} \\ & +p^{i_2}p^{i_3}\delta^{i_1i_4}+p^{i_1}p^{i_4}\delta^{i_2i_3}+p^{i_1}p^{i_3}\delta^{i_2i_4}
+p^{i_1}p^{i_2}\delta^{i_3i_4}] \\ & + [K_7 (p^0)^5 + K_8 (p^0)^4 + K_9 (p^0)^3 + K_{10}(p^0)^2 \\ & +K_{11} p^0 +K_{12}]\delta^{i_1i_2i_3i_4}.
\end{flalign*}
For each order $N$ of the polynomials, there are two groups of components: $P^{i_1\ldots i_N}$ and $P^{i_1\ldots i_{N-1}0}$. This structure assures that all possible monomials are considered for a generic spatial dimension $D$. The coefficients ($A$'s, $B$'s, $C$'s, $\ldots$) are calculated through the orthonormality relations:
\begin{flalign}\label{ortho-eq}
&\int \frac{d^D p }{p^{0}}\omega(p) P^{i_1\ldots i_N}P^{j_1\ldots j_M} = \delta_{NM} \delta^{i_1\ldots i_N|j_1\ldots j_N},\nonumber \\
&\int \frac{d^D p }{p^{0}}\omega(p) P^{i_1\ldots i_N0}P^{j_1\ldots j_M0} = \delta_{NM} \delta^{i_1\ldots i_N|j_1\ldots j_N},\nonumber \\ 
&\int \frac{d^D p }{p^{0}}\omega(p) P^{i_1\ldots i_N0}P^{j_1\ldots j_M} = 0.
\end{flalign}
Here, the normalization factor is the same as for the Hermite polynomials in D-dimensions~\cite{coelho14, doria17}, where we define $\delta_{i_1\cdots i_N\vert j_1\cdots j_N} \equiv \delta_{i_1 j_1}\cdots \delta_{i_Nj_N}\,+$ all permutations of  $j$'s and $\delta^{ij}$ is the Kronecker's delta. The weight functions used to build the models in this paper will be discussed in the next section (Eqs. \eqref{mj-w-eq}, \eqref{fd-w-eq} and \eqref{be-w-eq}) and the polynomial coefficients can be found in the Supplemental Material.

\subsection{Expansion}\label{expansion-sec}

In this section, we describe the fifth order expansion in relativistic polynomials of the MJ, BE and FD using a different set of polynomials for each distribution. Lets write the three EDFs in a general form:
\begin{flalign}\label{edf-general}
f^{eq} = \frac{A}{z^{-1} \exp \left(\frac{ \bar p_\alpha U^\alpha}{k_B T} \right) + \xi}
\end{flalign}
where A is a normalization factor, $z=e^{\frac{\bar\mu}{k_B T}}$ is the fugacity, $\bar\mu$ is the chemical potential and $\xi$ distinguishes the EDFs: $\xi=0$ for MJ (also $z=1$), $\xi=1$ for FD and $\xi=-1$ for BE. To expand the EDFs, we introduce non-dimensional quantities: $\theta = T/T_0$, $\mathbf{p}=\mathbf{\bar p}/T_0$ and $\mu = \bar \mu/T_0$, where $T_0$ is the reference temperature, e.g, the initial one. Thus, considering the ultra-relativistic regime, Eq. \eqref{edf-general} becomes
\begin{flalign}
f^{eq} = \frac{A}{z^{-1}\exp\left[ p^0 \gamma (1-\mathbf{\mathbf{v}}\cdot \mathbf{u} ) / \theta  \right] + \xi}.
\end{flalign}
As discussed in the previous section, the weight function must be close to the EDF. The expansion is done around $U=0$ (small Mach numbers) and around $\theta=1$, leading to the general weight function $\omega(p) = A/[z_r^{-1}\exp (p)+\xi ]$. The reference fugacity $z_r$ is a constant parameter with numerical value close to the physical one. Therefore, for the MJ distribution the weight function reads
\begin{flalign}\label{mj-w-eq}
\omega(p) = \frac{1}{2 \pi} e^{-p}.
\end{flalign}
For the FD distribution, we use $z=1$, which is appropriate, for instance, to model the Dirac fluid on graphene close to the charge neutrality point ($\mu=0$)~\cite{coelho17, furtmaier15}. So the weight function becomes
\begin{flalign}\label{fd-w-eq}
\omega(p) = \frac{1}{e^{p}+1}.
\end{flalign}
The weight function for the BE distribution is chosen to describe a system close to the Bose-Einstein condensation (ideally, $z=1$). Because some integrals diverge for $z_r=1$, we set $z_r \equiv z_c = 1 - \epsilon$, where $\epsilon = 10^{-7}$ denotes a small number compared to unity, leading to 
\begin{flalign}\label{be-w-eq}
\omega(p) = \frac{1}{z^{-1}_{c}e^{p}-1}.
\end{flalign}

\begin{figure*}[htb]
\centerline{\includegraphics[width=0.8\linewidth]{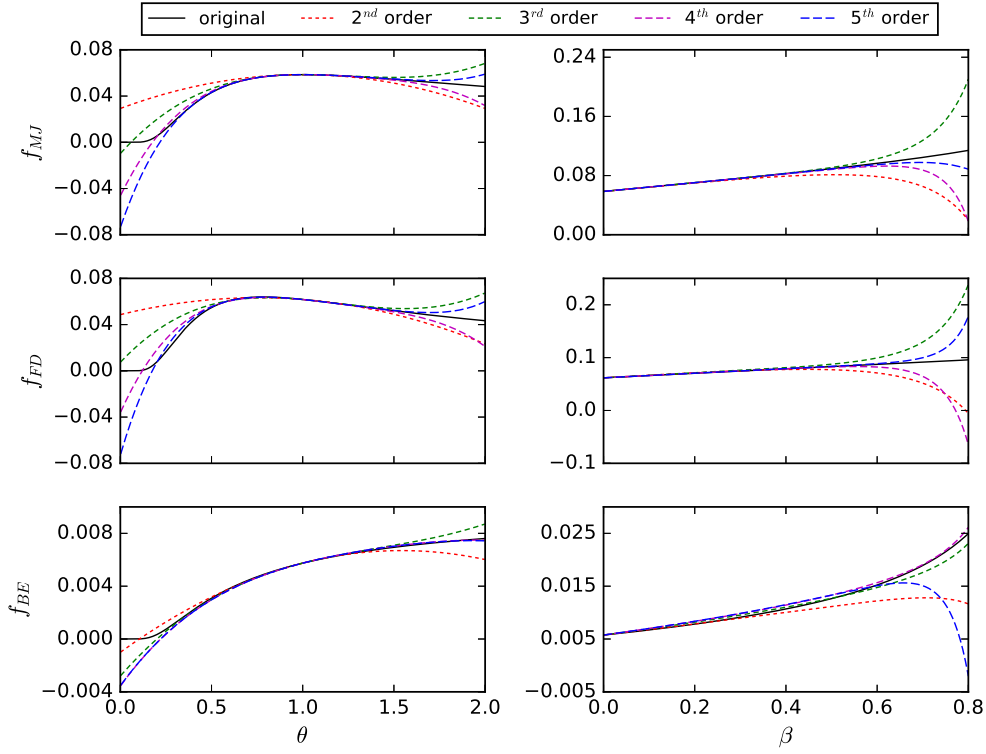}}
\vspace*{8pt}
\caption{Comparison between the original and expanded MJ, FD and BE distributions around the expansion origin: $\theta=1$ and $u=0$. The momentum vector is the same for all distributions: $\mathbf{p} = (1.0,\, 0.0)$. For FD, $z=1$ and for BE, $z=1-10^{-7}$. On the left side, the distributions are shown as functions of the relative temperature, $\theta = T/T_0$ (where $T$ and $T_0$ are the physical and the reference temperature respectively) with $\mathbf{u}=0$. On the right side, the EDFs are shown for different velocities, $\mathbf{u}/c=(\beta,\,0.0)$, with $\theta=1$.}
\label{comp-fig}
\end{figure*}

The general expansion up to fifth order is given by
\begin{flalign}\label{expansion-eq}
f^{eq} &= \omega(\xi)\left[ \sum ^5_{N=0}\frac{1}{N!} A^{i_1\ldots i_N}P^{i_1\ldots i_N} \right. \\ \nonumber  & \left. + \sum ^4_{M=0}\frac{1}{M!} A^{i_1\ldots i_M0}P^{i_1\ldots i_M0}   \right],
\end{flalign}
where $A^{\mu_1\,\mu_2 \cdots \mu_N}$ are the projections of the EDF on the polynomials:
\begin{flalign}\label{projections-eq}
A^{\mu_1\,\mu_2 \cdots \mu_N}= \int \frac{d^2 \mathbf{p} }{p^0}\,f^{eq} P^{\mu_1\,\mu_2\cdots \mu_N}.
\end{flalign}
More details about the calculation of the integrals used for the expansion can be found in the Appendix \ref{integrals-sec} and the explicit expansion can be found in the Supplemental Material. It was used in Ref.~\cite{coelho17} for FD distribution, but can be straightforwardly generalized to MJ and BE if one changes the polynomial coefficients and the FD integral by the generalized EDF integral defined as:
\begin{flalign}\label{gz-eq}
g_\nu(z)= \frac{1}{\Gamma(\nu)}\int_0^\infty \frac{x^{\nu-1}dx}{z^{-1}e^x+\xi},
\end{flalign}
with $\xi$ defined as in Eq.\eqref{edf-general}. Note that for MJ, $\xi=0$ and $g_\nu(1)=1$ for any $\nu$. In Fig. \ref{comp-fig}, we see the comparison between the original (non-expanded) EDFs and the expanded ones around the origin of the expansion ($\theta=1$ and $u=0$). The distributions are multiplied by a normalization factor $A=\left(  \int d^2 \mathbf{p} /p^0\,f^{eq} \right)^{-1}$ . The second order expansion is the first to deviate from the original EDF when it moves away from $\theta=1$ and $u=0$, followed by the third, fourth and fifth order expansions respectively. We can clearly see that the fifth order expansion gives more accurate results for higher temperature deviations and higher velocities when compared to the previous orders. For instance, considering the MJ distribution at $\beta=0.6$, the second order expansion gives a relative error when compared with the original EDF of 18.5\% while the fifth order gives only 1.8\%. At the extremes of Fig. \ref{comp-fig} ($\theta = 0\mbox{ and } 2$ and $\beta=0.8$) we see large deviations for all expansion orders, but one should not expect high accuracy for parameters very far from the expansion origin. Therefore, the models based on the fifth order expansion are more reliable and can be used to simulate flows with higher Mach numbers and temperature deviations than the usual models based on the second order expansion. In Sec. \ref{hot-spot-sec}, we will analyze the different results in numerical simulations based on expansions from second to fifth order.

\subsection{Gaussian Quadrature}\label{quadrature-sec}

\begin{figure}[htb]
\centerline{\includegraphics[width=0.9\linewidth]{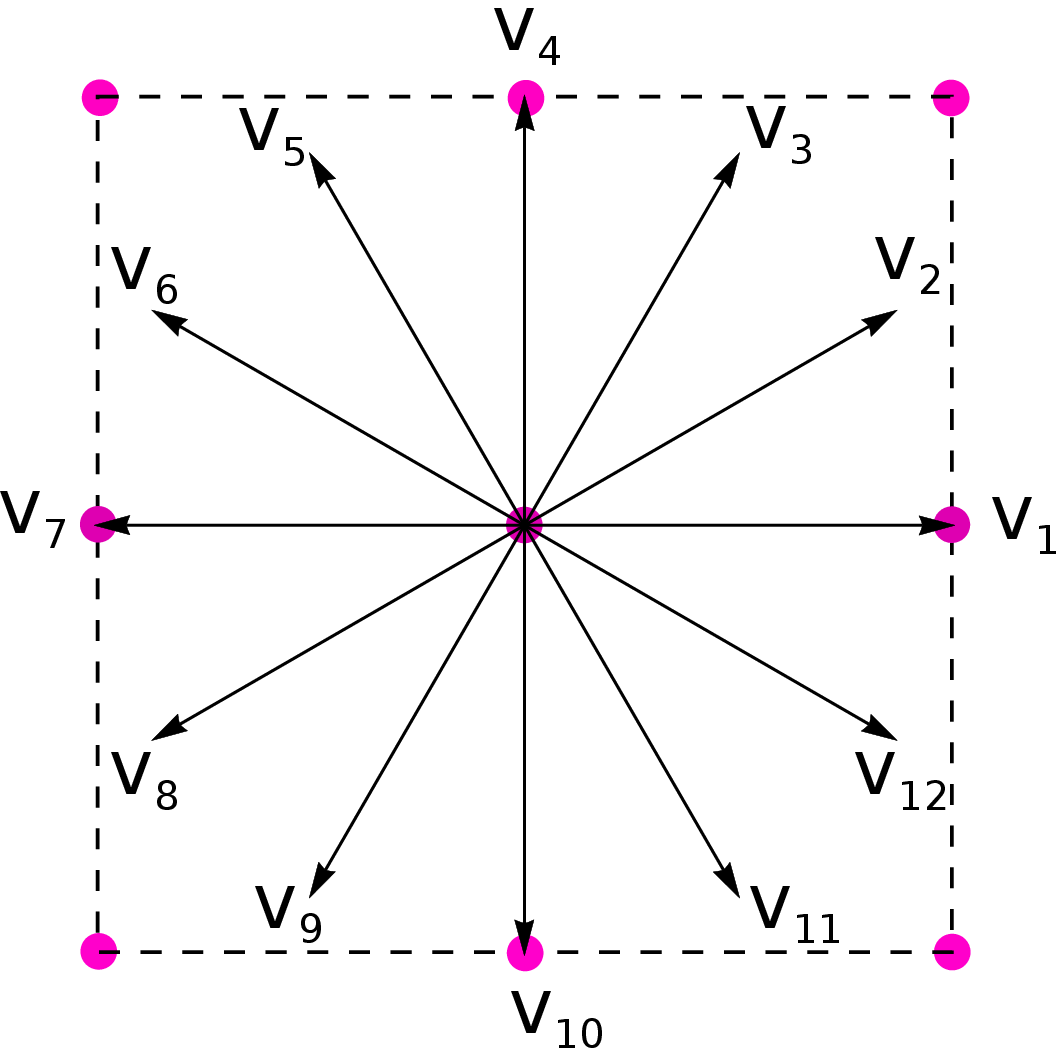}}
\vspace*{8pt}
\caption{Velocity vectors for the d2v72 quadrature.}
\label{d2v72}
\end{figure}
The Gaussian quadrature~\cite{abramowitz64} offers an efficient way to calculate integrals in LBM, used to obtain the macroscopic quantities (e.g, density, macroscopic velocity and temperature), by transforming them into sums. In general, the method provides an approximation for the integrals, but it can be exact if the integrated function is expanded in orthogonal polynomials up to a maximum order determined by the quadrature. Here, we calculate quadratures that allow us to calculate tensors up to fifth order ($M=5$),
\begin{flalign*}
T^{\mu_1  \ldots \mu_M} &= \int \frac{d^2 p}{p^0} f^{eq} p^{\mu_1} \ldots p^{\mu_M} \\&= \sum_{i=1}^Q f^{eq}_i p^{\mu_1}_i \ldots p^{\mu_M}_i .
\end{flalign*}
Since we have a fifth order expansion for our EDFs, we need to find the discrete weights and momentum vectors that satisfy the quadrature equations,
\begin{flalign}\label{quadrature-eq}
\int \frac{d^2 p}{p^0} \omega(p) p^{\mu_1}  \ldots p^{\mu_N} = \sum_{i=1}^Q w_i p^{\mu_1}_i \ldots p^{\mu_N}_i,
\end{flalign}
for $N=0,1,\ldots,10$ and for all combinations of indexes ($\mu=0, \, 1, \, 2$). The weight function, $\omega(p)$, used to calculate the quadrature and the polynomials must be the same for each model (Eqs. \eqref{mj-w-eq}, \eqref{fd-w-eq}, \eqref{be-w-eq}). As a consequence of the quadrature theorem, we can find the momentum vectors for our quadrature (with $N=10$) by calculating the roots of the sixth order polynomial, which can be separated in angular and radial parts. For the angular part, there are two orthogonal polynomials,
\begin{flalign*}
L_6^{(1)}(\phi)=\cos (6 \phi ) ,\: L_6^{(2)}(\phi)=\sin (6 \phi ) ,
\end{flalign*}
which give the same set of solutions, but rotated by $\pi/12$. The roots of $L_6^{(2)}(\phi)=0$, are $\phi_n=n\pi/6$ for $n=0,\ldots,11$, which are the directions of the momentum vectors (see Fig. \ref{d2v72}). The sixth order polynomial for the radial part,
\begin{flalign*}
 R_6(p)= p^6+ap^5+bp^4+cp^3+dp^2+ep+f,
\end{flalign*}
is calculated by Gram-Schmidt procedure, where the coefficients, $a,\ldots,f$, are different for each weight function. The solutions of $R_6(p)=0$ give six momentum vectors, which are the same for each of the 12 directions obtained with the angular polynomial. Therefore the total number of momentum vectors of our quadrature is 72 (d2v72 quadrature). To calculate the discrete weights, we apply Eq.\eqref{quadrature-eq}. The solutions for MJ, using Eq.\eqref{mj-w-eq}, are
\begin{flalign*}
p_1 = 0.222847 \:\:\:\:\:\:\:\:& w_1 = 3.824706\times 10^{-2}\\ 
p_2 = 1.188932\:\:\:\:\:\:\:\: & w_2 = 3.475007\times 10^{-2} \\
p_3 = 2.992736\:\:\:\:\:\:\:\: & w_3 = 9.447782 \times 10^{-3}\\
p_4 = 5.775144\:\:\:\:\:\:\:\: & w_4 = 8.665998\times 10^{-4}\\
p_5 = 9.837467 \:\:\:\:\:\:\:\:& w_5 = 2.175143\times 10^{-5}\\
p_6 = 15.982874\:\:\:\:\:\: & w_6 = 7.487899\times 10^{-8},
\end{flalign*}
and for BE, using Eq.\eqref{be-w-eq}, 
\begin{flalign*}
p_1 = 0.015771 \:\:\:\:\:\:\:\:&w_1 =  7.785968\\ 
p_2 = 0.811549 \:\:\:\:\:\:\:\:&w_2 =  5.510286\times 10^{-1} \\
p_3 = 2.617676\:\:\:\:\:\:\:\: & w_3 = 9.444240\times 10^{-2}\\
p_4 = 5.428098 \:\:\:\:\:\:\:\:& w_4 = 7.784734\times 10^{-3}\\
p_5 = 9.503366 \:\:\:\:\:\:\:\: & w_5 =1.911972\times 10^{-4}\\
p_6 = 15.65413 \:\:\:\:\:\:\:\:& w_6 = 6.539175\times 10^{-7}.
\end{flalign*}
For FD, using Eq.\eqref{fd-w-eq}, the quadrature was calculated in Ref.~\cite{coelho17}. The three quadratures can be found with higher precision in the Supplemental Material.

Since the lattice vectors do not match the cartesian grid, the distribution functions at the grid points are found by means of interpolations, which are bilinear in our models. In Fig. \ref{d2v72}, we see that the directions that needs interpolation are $i=2,\,3,\,5,\,6,\,8,\,9,\,11,\,12$. Thus, to find the value of the distribution function propagated in one of these directions the following equation apply:
\begin{eqnarray}
&&f_i(x,y, t) = f_i(\bar x_i,\bar y_i, t-\delta t)\vert \text{v}_i^{(x)} \vert \vert \text{v}_i^{(y)}\vert \nonumber\\
&&  
+f_i(x,\bar y_i, t-\delta t)(1-\vert \text{v}_i^{(x)} \vert) \vert \text{v}_i^{(y)}\vert \nonumber\\
&&+ 
f_i(\bar x_i,y, t-\delta t)\vert \text{v}_i^{(x)} \vert(1- \vert \text{v}_i^{(y)}\vert) \nonumber\\
&&+
f_i(x,y,t-\delta t)(1-\vert \text{v}_i^{(x)} \vert)(1- \vert \text{v}_i^{(y)}\vert), 
\end{eqnarray}
where $\bar x_i$ and $\bar y_i$ are the previous positions ($\bar x_i = x - \text{v}^{(x)}_i$ and $\bar y_i = y - \text{v}^{(y)}_i$ for the bulk) and $\mathbf{v_i} = \mathbf{p}_i/ \vert \mathbf{p}_i\vert$. For the other directions, $i=1,\,4,\,7,\,10$, the streaming is as usual:
\begin{eqnarray}
 f_i(x,y,t) = f_i(\bar x, \bar y, t-\delta t).
\end{eqnarray}

To calculate the macroscopic fields, we use the Landau-Lifshitz decomposition~\cite{cercignani02}. We first calculate the energy-momentum tensor
\begin{flalign}
T^{\mu\nu} = \sum_{i=1}^Qf_i p^\mu_i p^\nu_i,
\end{flalign}
and, then, solve the following eigenvalue problem using the power method~\cite{hoffman2001numerical}:
\begin{flalign}\label{eigen-eq}
{T_E^{\alpha}}_\beta U^\beta = {T^{\alpha}}_\beta U^\beta =  \varepsilon U^\alpha
\end{flalign}
to find the energy density $\varepsilon$ and the macroscopic velocity $U^\mu$, where the letter $E$ indicates that the tensor was calculated with the equilibrium distribution. The hydrostatic pressure $P$ can be obtained using the equation of state, $\varepsilon = 2P$. The density of particles, $n$, is calculated as the contraction of the macroscopic velocity with the particles flux $N^\mu$,
\begin{flalign}\label{density-eq}
n =  U_\mu N_E^\mu = U_\mu N^\mu = U_\mu\sum_{i=1}^Qf_i p^\mu_i.
\end{flalign}
Lastly, the temperature is calculated with
\begin{flalign}\label{theta-eq}
\theta = \frac{1}{2} \frac{g_2(z)}{g_3(z)}\left(\frac{\varepsilon}{n}\right),
\end{flalign}
where the EDF integral, $g_\nu(z)$, is defined in Eq.\eqref{gz-eq}. 

The density of particles and the internal energy can be calculated with the equilibrium distribution, since, by Eqs.\eqref{eigen-eq} and \eqref{density-eq}, they give the same result as for the non-equilibrium distribution:
\begin{flalign}\label{n-eps-eq}
n = 2\pi\theta^2 g_2(z),\:\: \mbox{and}\:\: \varepsilon = 2P = 4 \pi \theta^3 g_3(z).
\end{flalign}
The Eq.\eqref{theta-eq} was calculated using Eq.\eqref{n-eps-eq}. The validity of  Eqs.\eqref{eigen-eq} and \eqref{density-eq} is a consequence of the conservation of particles flow and the conservation of the energy-momentum tensor,
\begin{flalign}\label{conserv-eq}
\partial _\mu N^\mu = 0,\:\:\:\: \partial _\mu T^{\mu \nu} =0.
\end{flalign}
To calculate the transport coefficients using the Grad's expansion, one also needs an equation for the third order non-equilibrium tensor~\cite{mendoza13-3}, which requires the fifth order equilibrium tensor, 
\begin{flalign}\label{fifth-tensor-eq}
T_E^{\alpha \beta \gamma \delta \epsilon} = \int f^{eq} p^\alpha p^\beta p^\gamma p^\delta p^\epsilon \frac{d^2p}{p^0},
\end{flalign}
so, an expansion up to, at least, fifth order is needed to recover the full dissipation. Another way to see the required expansion order of the EDF, is to observe that the pressure deviator, Eq.\eqref{press-dev-eq}, has terms with five velocities, which can be recovered just with a fifth order expansion.
In the Landau-Lifshitz decomposition, the particle flow reads~\cite{cercignani02}
\begin{flalign}\label{charge-flow-eq}
N^\mu = n U^\mu - \frac{q^\mu}{h_E},
\end{flalign}
where $q^\mu$ is the heat flux,
\begin{flalign}\label{heat-flux-eq}
q^\mu = \kappa \left(\nabla^\mu T - \frac{T}{c^2}D U^\mu \right),
\end{flalign}
 $\kappa$ is the thermal conductivity, $D= U^\alpha \partial_\alpha$ and $h_E = (\varepsilon + P)/n = 3\,T\,g_3(z)/g_2(z)$ is the enthalpy per particle, and the energy-momentum tensor is written as
\begin{flalign}
T^{\mu\nu} = p^{\langle \mu \nu  \rangle} - (P + \varpi)\Delta^{\mu\nu} + \frac{\varepsilon}{c} U^\mu U^\nu,
\end{flalign}
where 
\begin{flalign}\label{press-dev-eq}
& p^{\langle \mu \nu \rangle} = 2\eta \left [ \frac{1}{2}(\Delta^\mu_\gamma \Delta^\nu_\delta + \Delta^\mu _\delta \Delta ^\nu _\gamma ) - \frac{1}{3}\Delta ^{\mu\nu} \Delta_{\gamma \delta} \right] \nabla^\gamma U^{\delta}\nonumber \\
&
\end{flalign}
is the pressure deviator, $\varpi = - \mu_b \nabla_\alpha U^\alpha$ is the dynamic pressure, $\eta$ is the shear viscosity and $\mu_b$ is the bulk viscosity. Here, $\Delta^{\mu \nu} = \eta^{\mu \nu} - U^\mu U^\nu/c^2$ stands for the projector into the space perpendicular to $U^\mu$ and $\nabla^\mu = \Delta^{\mu \nu} \partial_\nu$ for the gradient operator. 

In the next section, we adopt a reference model (denoted as ``ref.'' in the figures) to validate our new models and compare our results with the two-dimensional version of the LBM described in Ref.~\cite{mendoza13-3}. This model is based on a third order expansion, and therefore less accurate than the models presented here, but it has the advantage to use a quadrature with exact streaming, which allows us to analyze the effects of our interpolated streaming. Because all ultra-relativistic particles move essentially with the same speed, to find a quadrature with exact streaming on a square lattice, one has to find velocity vectors belonging, simultaneously, to the lattice nodes and to a circle of radius $R$:
\begin{flalign*}
n_x^2+n_y^2=R^2,
\end{flalign*} 
where $n_x$ and $n_y$ are integers. The simplest lattice that satisfies the above equation and the quadrature equations, Eq. \eqref{quadrature-eq}, up to $N=6$ (third order model) has radius $R=5$ and 36 momentum vectors, where the velocity vectors are:
\begin{flalign*}
 &v_x = \{-5, -4, -4, -3, -3, 0, 0, 3, 3, 4, 4, 5\}\\
 &v_y = \{0, -3, 3, -4, 4, -5, 5, -4, 4, -3, 3, 0\}
\end{flalign*}
The main disadvantage of this quadrature is a loss of resolution. As a consequence, to simulate the same two-dimensional physical system one needs $5\times5=25$ times more lattice points and therefore more memory and time. In addition, this quadrature can not be used to simulate models based on expansions of orders higher than three, like our fifth order models. This would require a much bigger radius $R$, making the resolution so small that the model would have little practical utility. So, the interpolated streaming arises as an alternative to keep the resolution and to have higher order expansions.

\section{Numerical validation and characterization}\label{numerical-sec}

\subsection{Riemann Problem}\label{riemann-sec}

\begin{figure}[htb]
\centerline{\includegraphics[width=1.0\linewidth]{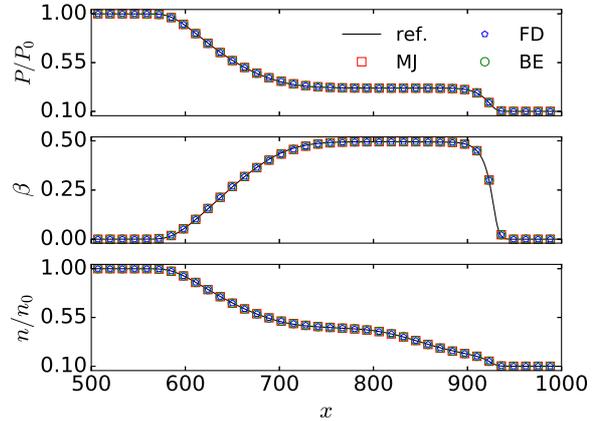}}
\vspace*{8pt}
\caption{Solution of the Riemann problem after $t=200$ time steps for the pressure, velocity and density fields using our three models and the reference model. The pressure and the density fields are divided by their initial value at the center ($x=500$).}
\label{shocktube-fig}
\end{figure}

The Riemann problem consists of a discontinuity in the initial conditions of the macroscopic quantities (e.g., density or velocity) which generates shock waves when the system evolves. This is a benchmark validation for fluid dynamics models which has analytic solution for the inviscid hydrodynamic equations~\cite{toro09} but needs to be solved numerically for viscous fluids~\cite{bouras09}. In order to validate our models, we compare solutions for the pressure, velocity and density fields with a reference model described in Ref.~\cite{mendoza13-3}. Initially, we set the density $n_0=1.0$ in the domain $L_X/4 < x < 3L_X/4$ and $0.1$ elsewhere, with temperature $\theta_0=1.0$ and velocity $\mathbf{u_0}=0.0$ on the whole domain. The effectively one-dimensional system has dimensions $L_X \times L_Y = 1000 \times 2$ with periodic boundary conditions in both directions. Just half of the system is shown because the other half is an exact mirror image. The reference model, as well as the previous relativistic LBMs, uses the viscosity given by Grad's expansion, but, as we will see in Sec. \ref{viscosity-sec}, the measured viscosity differs from this theoretical prediction. Because of this discrepancy, instead of using constant $\eta/s$ as usual, we perform our simulation using a constant relaxation time of $\tau=10.0$ and, since the viscosity is constant, the results from the three EDFs are basically the same for the Riemann problem. In Secs. \ref{viscosity-sec} and \ref{thermal-sec} we will obtain the transport coefficients for each EDF and compare them with the ones from the literature. In Fig. \ref{shocktube-fig}, we see that the three models agree very well with the reference model for those fields. Notice that the pressure differs for the three EDFs due to the EDF integrals (see Eq.\eqref{n-eps-eq}), but, since it is scaled by the initial pressure at $x=500$, the curves coincide in Fig. \ref{shocktube-fig}.

\begin{figure*}[htb]
\centerline{\includegraphics[width=0.8\linewidth]{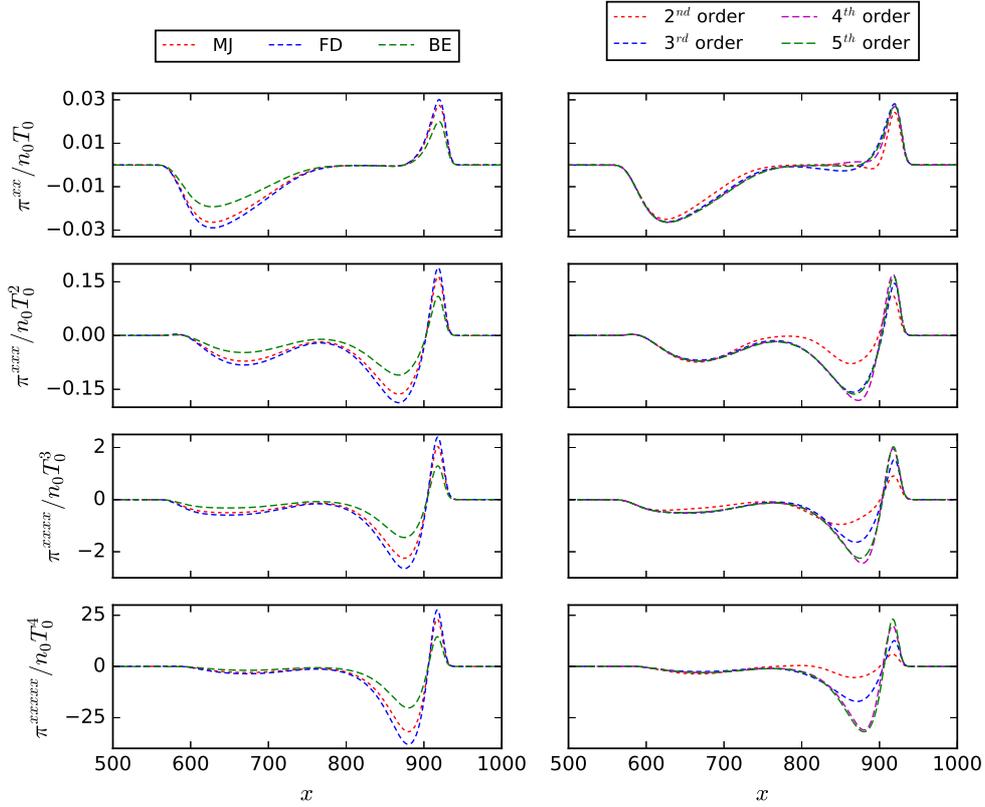}}
\vspace*{8pt}
\caption{The diagonal components of the second to fifth order tensors (see Eq.\eqref{tensors-eq}) calculated using the three EDFs expanded up to fifth order (left side) and with the MJ distribution expanded from second to fifth order in orthogonal polynomials (right side). The tensors are divided by the initial density at the center ($x=500$), $n_0$, and by the initial temperature $T_0$ to remove the units.}
\label{tensors-fig}
\end{figure*}

In order to show the differences between different expansion orders of the EDF, we calculate the diagonal components of the tensorial fields, from second to fifth order, in the Riemann problem:
\begin{flalign}\label{tensors-eq}
 &\pi^{xx} = \sum _i (f-f^{eq}) p^x p^x,\\ \nonumber & \pi^{xxx} = \sum _i (f-f^{eq}) p^x p^x p^x,\\ \nonumber
 &\pi^{xxxx} = \sum _i (f-f^{eq}) p^x p^x p^x p^x,\\ \nonumber &  \pi^{xxxxx} = \sum _i (f-f^{eq}) p^x p^x p^x p^x p^x.
\end{flalign}
The results can be seen in Fig. \ref{tensors-fig} for the three EDFs expanded up to fifth order on the left and for the MJ distribution expanded from second to fifth order on the right. The numerical values of the tensors are divided by the initial density and temperature at the center ($x=500$) in order to make them non-dimensional. The differences between the three EDFs clearly appear for these tensors. Considering the MJ distribution with different expansion order (on the right), we see that all models give similar results for the second order tensor, but the differences increase with the order of the tensors, becoming large for the fifth order tensor. As in Fig. \ref{comp-fig}, the results seem to converge when the expansion order increases. This shows the importance of using higher order expansions when higher order tensors are considered.

\subsection{Viscosity measurement}\label{viscosity-sec}

\begin{figure*}[htb]
\centerline{\includegraphics[width=0.8\linewidth]{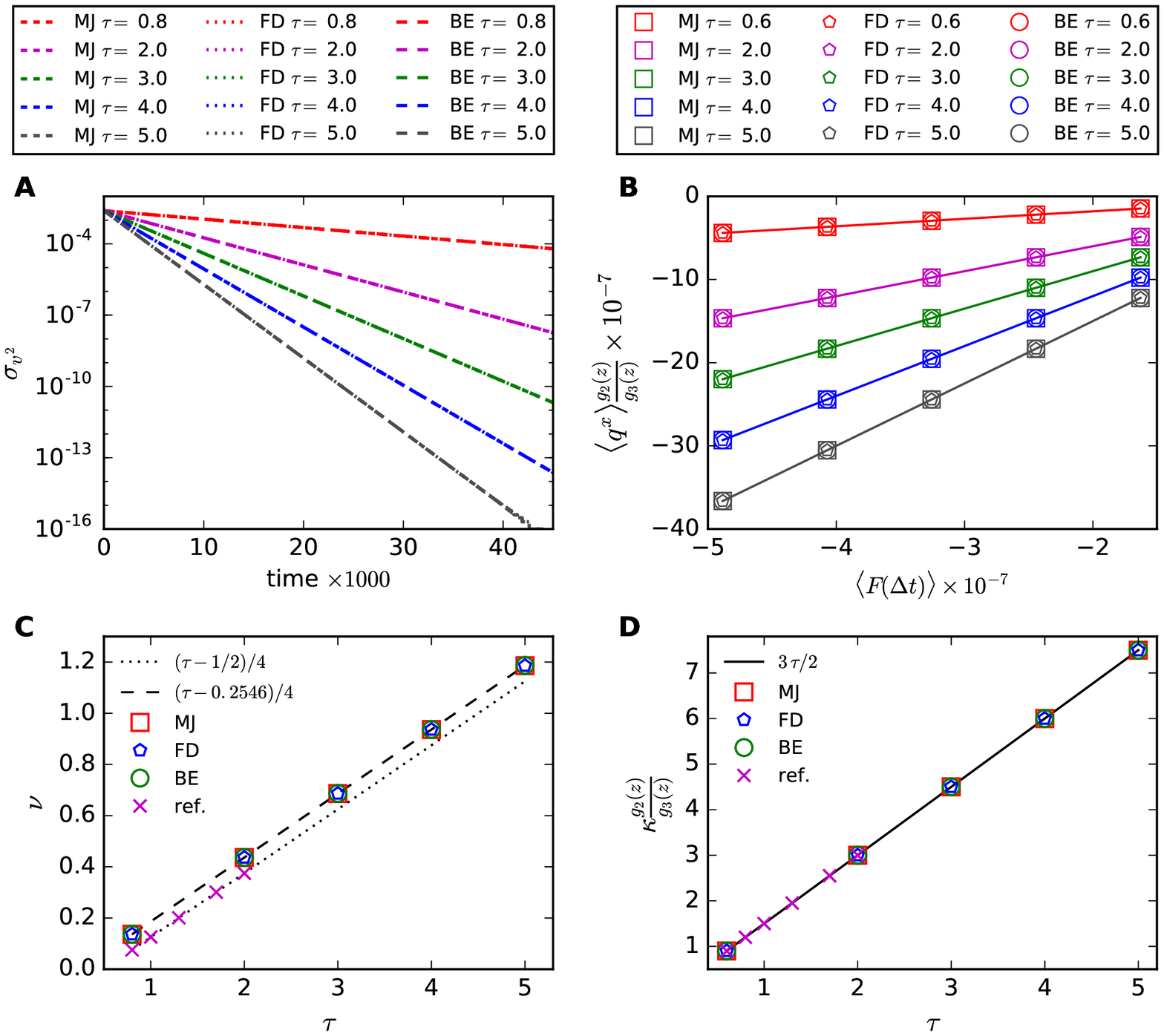}}
\vspace*{8pt}
\caption{Measurement of transport coefficients. A) Decay with time of $\sigma_{u^2}$ for the three distributions and five different relaxation times $\tau$ in the Taylor-Green vortex. The curves for the same $\tau$ fall on top of each other. B) Average heat flux as a function of $\langle F(\Delta T) \rangle$ (Eq.\eqref{fdt-eq}) for the three distributions and five relaxation times. The solid lines represent linear fits, which fall on top of each other for the three distributions. C) Kinematic viscosity -- relaxation time relation for the three distributions obtained with the Taylor-Green vortex and comparison with the results from the reference model. D) Thermal conductivity-- relaxation time relation for the three distributions and comparison with the reference model. }
\label{eta-kappa-fig}
\end{figure*}

We measure the kinematic viscosity in our models through the Taylor-Green vortex experiment. This is a initial value problem consisting of initial vortexes rotating in determined directions, which kinetic energy dissipates with time due to the viscosity. The conservation equations, Eq.\eqref{conserv-eq}, can be solved exactly for this problem if one considers low velocities compared to the speed of light, giving exponential decay with time for the velocity field depending on the kinematic viscosity $\nu$~\cite{gabbana17, mei06}: 
\begin{flalign}
\mathbf{u}(x,y,t) = \mathbf{u}_{0}(x,y)e^{-2\nu t (2\pi/L)^2},
\end{flalign}
where $\mathbf{u}_0$ is the initial velocity and $L$ the length of the squared domain. We simulate a system with dimensions $L\equiv L_X=L_Y=512$ and  with periodic boundary conditions for five different relaxation times, ranging from 0.8 to 5.0 for 45000 time steps. The initial conditions are $n_0=1.0$ and $\theta _0=1.0$ in the whole domain and the initial velocities are:
\begin{flalign}
&u_{0x} (x,y) = - u_0  \cos\left(\frac{2\pi x}{L}\right)  \sin\left(\frac{2\pi y}{L}\right)   \\ 
&u_{0y} (x,y) =  u_0  \sin\left(\frac{2 \pi x}{L}\right)  \cos\left(\frac{2 \pi  y}{L}\right)  , 
\end{flalign}
where $u_0=0.1$. The initial non-equilibrium distribution is set as described in Ref.~\cite{mei06} in order to reduce the oscillations in the fields. So the average squared velocity writes
\begin{flalign}
\langle u^2 \rangle = \int_0^{L}\int_0^{L} \frac{dx dy}{L^2}(u_x^2+u_y^2) = \frac{u_0^2}{2}e^{-16\pi^2\nu t/L^2} ,
\end{flalign}
and the standard deviation for $u^2$ is given by
\begin{flalign}\label{sigma-u2}
\sigma_{u^2} = \sqrt{\int _0 ^L \int _0^L \frac{dx dy}{L^2} (u^2 - \langle u^2 \rangle )^2}  =\frac{u_0^2}{4} e^{-16\pi^2\nu t/L^2}  
\end{flalign}
In Fig. \ref{eta-kappa-fig}A we see $ \sigma_{u^2} $ as a function of time in semi-log scale for the three models. From the slope of $ \sigma_{u^2}(t)$, we can measure the kinematic viscosity $\nu$. Note that $\nu$ does not depend on the distribution, but the shear viscosity, $\eta = (\varepsilon + P)\nu$, does. Fig. \ref{eta-kappa-fig}C shows the measured kinematic viscosity as a function of the relaxation time.
The relation
\begin{flalign}\label{fiducial-visc-eq}
\nu(\tau) = \frac{1}{4} \left( \tau - \frac{\delta t}{2} \right),
\end{flalign}
shows good agreement for ultra-relativistic models based on exact streaming~\cite{furtmaier15}, but the interpolated streaming introduces a numerical
diffusivity which increases the effective viscosity of the fluid~\cite{PhysRevE.61.6546, Wu20112246, Yu2003329}, i.e.,
\begin{flalign}
\nu_{eff} = \frac{1}{4}\left[ \tau  - \delta t\left(\frac{1}{2} + \delta_\nu \right) \right] .\label{nu-ref-eq}
\end{flalign}
With a linear fit, we measure the increment $\delta_\nu$ in the viscosity for the three EDFs:  
\begin{flalign*}
&\delta_\nu ^{MJ}= -0.2454 \pm 0.0001 \\ 
& \delta_\nu ^{FD}= -0.2454 \pm 0.0002 \\
& \delta_\nu ^{BE}= -0.2449 \pm 0.0005.
\end{flalign*}
These results can be summarized as $\nu_{eff} = \frac{1}{4}\left( \tau  - 0.2546\,\delta t  \right)$; see in Fig. \ref{eta-kappa-fig}C this function compared to the data from the simulations. In order to have a more realistic thermodynamic behavior, the shear viscosity-entropy ratio ($\eta/s$) is set constant and the relaxation time is calculated with $\tau = 4 \eta/(s \theta) + 0.2546\,\delta t $, which will be used in Sec. \ref{hot-spot-sec}.

In order to compare our results with the viscosity from a model with exact streaming, we perform the same numerical experiment with the reference model, but with different system dimensions, $L_X \times L_Y = 320\times 320$, due to the differences in the resolution described in Sec. \ref{quadrature-sec}. Notice that this system size would be equivalent to $L_X \times L_Y = 64 \times 64$ for the models described in this paper. A linear fit $\nu(\tau)=a(\tau-b)$ gives:
\begin{flalign*}
& a = 0.2502 \pm 0.0003\\
& b = 0.4996 \pm 0.0005
\end{flalign*}
confirming Eq.\eqref{fiducial-visc-eq}, see Fig. \ref{eta-kappa-fig}C. Note that, although this result was obtained directly from the Boltzmann equation, it is not compatible with the prediction from Grad's expansion~\cite{mendoza13-2}, $\nu_{Grad} = k_B \tau /5$, what underlines the need for a better understanding about the transport coefficients of relativistic fluids in two dimensions. 

\subsection{Thermal Conductivity Measurement}\label{thermal-sec}

From the correspondence between the Eckart and Landau-Lifshitz decompositions~\cite{cercignani02}, one can calculate the heat flux directly by the macroscopic fields (see Eq.\eqref{charge-flow-eq}):
\begin{flalign}\label{heat-eq2}
q^\alpha = \frac{3T g_3(z)}{g_2(z)}( n U^\alpha - N^\alpha).
\end{flalign}
Thus, combining Eq.\eqref{heat-flux-eq} with Eq.\eqref{heat-eq2}, we can calculate the thermal conductivity of the fluid. Considering a one dimensional gradient in temperature in $x$-direction, Eq.\eqref{heat-flux-eq} becomes
\begin{flalign}\label{qx-ft-eq}
q^x =\kappa F(\Delta T) 
\end{flalign}
where
\begin{flalign}\label{fdt-eq}
& F(\Delta T) \equiv -  \left\{ \left( 1+\frac{(u^x)^2 \gamma^2}{c^2} \right) \frac{\partial T}{\partial x} \right. \nonumber \\ & \left. +\frac{T\gamma}{c^2}
\left[ c\frac{\partial}{\partial t}(\gamma u^x) + u^x \frac{\partial}{\partial x}(\gamma u^x)  \right]      \right\}.
\end{flalign}
In the non-relativistic limit, Eq.\eqref{qx-ft-eq} becomes Fourier's law, while $F(\Delta T) \rightarrow - \partial T/\partial x$. 

In order to measure the thermal conductivity we simulate an effectively one dimensional system with dimensions $L_x\times L_y = 2048\times 2$ for 5 different gradients in temperature in the $x$ direction.  We calculate the spatial average of $F$, $\langle  F(\Delta T) \rangle$, with Eq.\eqref{fdt-eq} and the average heat flux, $\langle q^x   \rangle$  with Eq.\eqref{heat-eq2}, where both are essentially
constant in space. For each simulation, the temperatures on the left and right boundaries are constant and set as $\theta_L = 1-\Delta T/2$ on the left and $\theta_R=1+\Delta T /2$ on the right, while the differences in temperature are
$\Delta T = \{5.0,\, 7.5,\, 10.0, \,12.5,\, 15.0\}\times 10^{-4}$. A zeroth order extrapolation using the first fluid neighbors is performed to find the density and velocity on left and right borders and periodic boundary condition are used on top and bottom. The initial conditions are $n_0=1.0$ and $\mathbf{u_0}=0$
everywhere and we set an initial temperature gradient as $\theta_0(x) =  \theta_L + x (\theta_R - \theta_L)/L_x$ to have a faster convergence to the solution. Fig. \ref{eta-kappa-fig}B shows the average heat flux as a function of $\langle F(\Delta T) \rangle $ for 5 relaxation times and for the three EDFs, and their respective linear fits (overlap for the three distributions) after 2000 time steps. The slope of each line gives the thermal conductivity,
which can be seen in Fig. \ref{eta-kappa-fig}D as a function of the relaxation time. The linear fits, $[\kappa g_2(z)/g_3(z)](\tau)=a\tau$, for the three EDFs give:
\begin{flalign*} 
&a_{MJ}=1.4999998 \pm 0.0000002\\
&a_{FD}=1.4999998 \pm 0.0000002\\
&a_{BE}=1.4999997 \pm 0.0000004, 
\end{flalign*}
suggesting that the thermal conductivity-relaxation time relation is 
\begin{flalign}\label{kappa-tau-eq}
\kappa(\tau) = \frac{3\,\tau \,g_3(z)}{2\, g_2(z)}.
\end{flalign}
This relation also agrees with the thermal conductivity obtained using the reference model (with system dimensions $L_X \times L_Y = 2560 \times 5 $), as shown in Fig. \ref{eta-kappa-fig}D, for which we find
\begin{flalign*}
a_{ref} =  1.4999994 \pm 0.0000008,
\end{flalign*}
with the linear fit. Interestingly, this results shows that the interpolated streaming changes the viscosity but not the thermal conductivity. A similar relation was found in Ref.~\cite{furtmaier15}, where the authors obtained $\kappa(\tau)=1.525 \tau$, with a relative error of $7.2\%$ compared to our result, possibly due to the small resolution of the numerical experiment ($L_x\times L_Y=32 \times 32$). Similarly for the viscosity in Sec. \ref{viscosity-sec}, the thermal conductivity obtained with relativistic LBM differs from the one predicted by Grad's expansion~\cite{mendoza13-2}, which is $\kappa_{Grad}=3c^2k_Bn\tau /8$ for MJ distribution. It shows the importance to perform a careful Chapman-Enskog expansion in two dimensions, since other works in three dimensions have shown very good agreement between LBM and Chapmann-Enskog~\cite{gabbana17}. 

In order to convert from lattice units to physical units it is necessary to know the transport coefficients of the specific fluid that is being simulated. Then, the non-dimensional numbers of interest (e.g, the Reynolds number: $Re = L \, u/\nu$, where $L$ is a representative length of the system, $u$ is the fluid velocity and $\nu$ is the kinematic viscosity) should be equal in both unity systems~\cite{landau86, kruger16}. Since in this paper we are interested in the numerical analysis of generic models, we will not convert to physical units for a specific problem. See, for instance, Refs.~\cite{coelho17, oettinger13, furtmaier15} for units conversion involving graphene problems (Fermi-Dirac distribution) and Refs.~\cite{mendoza10, mendoza13-3, hupp11} for problems involving quark-gluon plasma.

\subsection{Grid convergence}

\begin{figure}[htb]
\centerline{\includegraphics[width=1.0\linewidth]{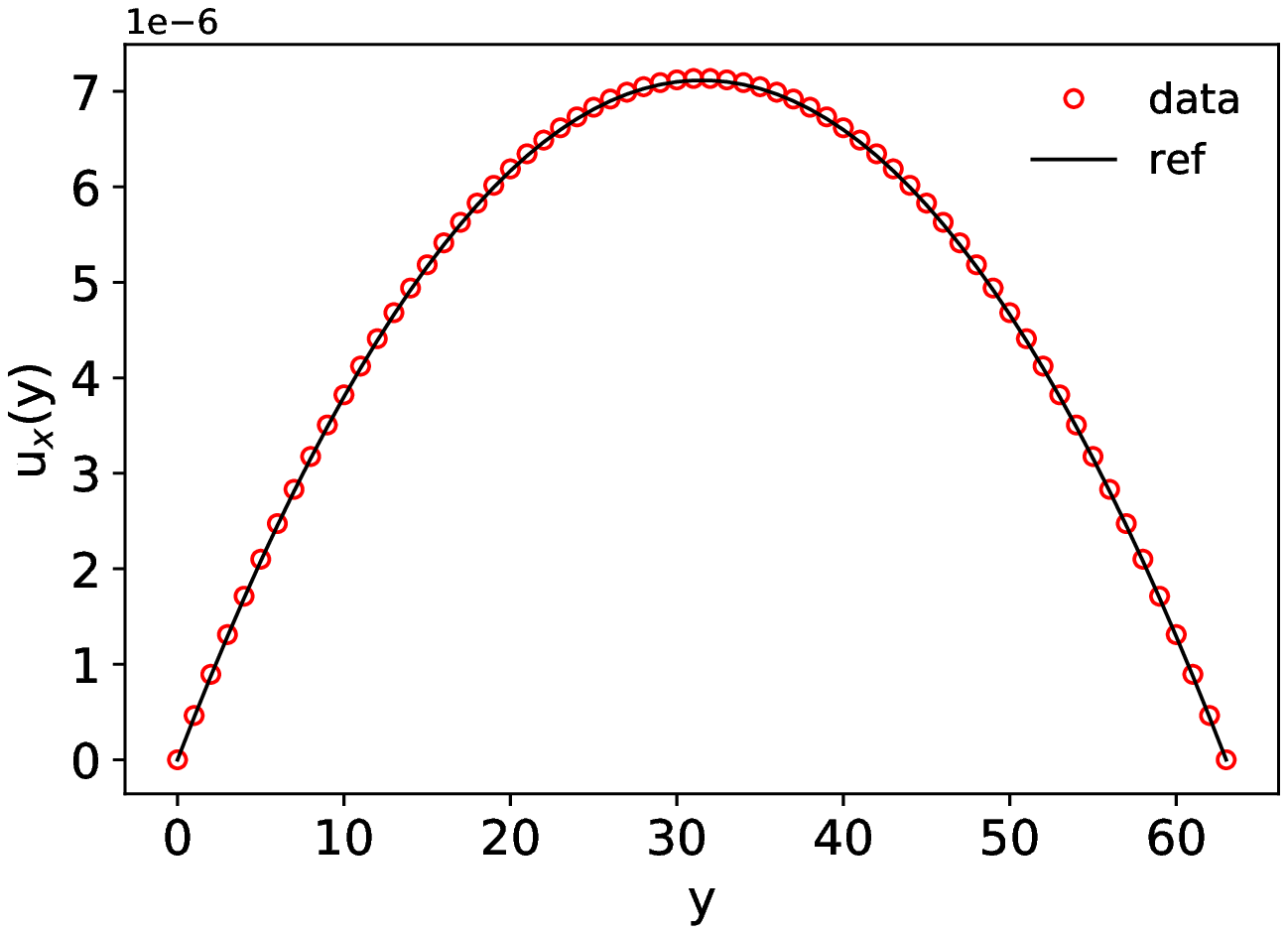}}
\centerline{\includegraphics[width=1.0\linewidth]{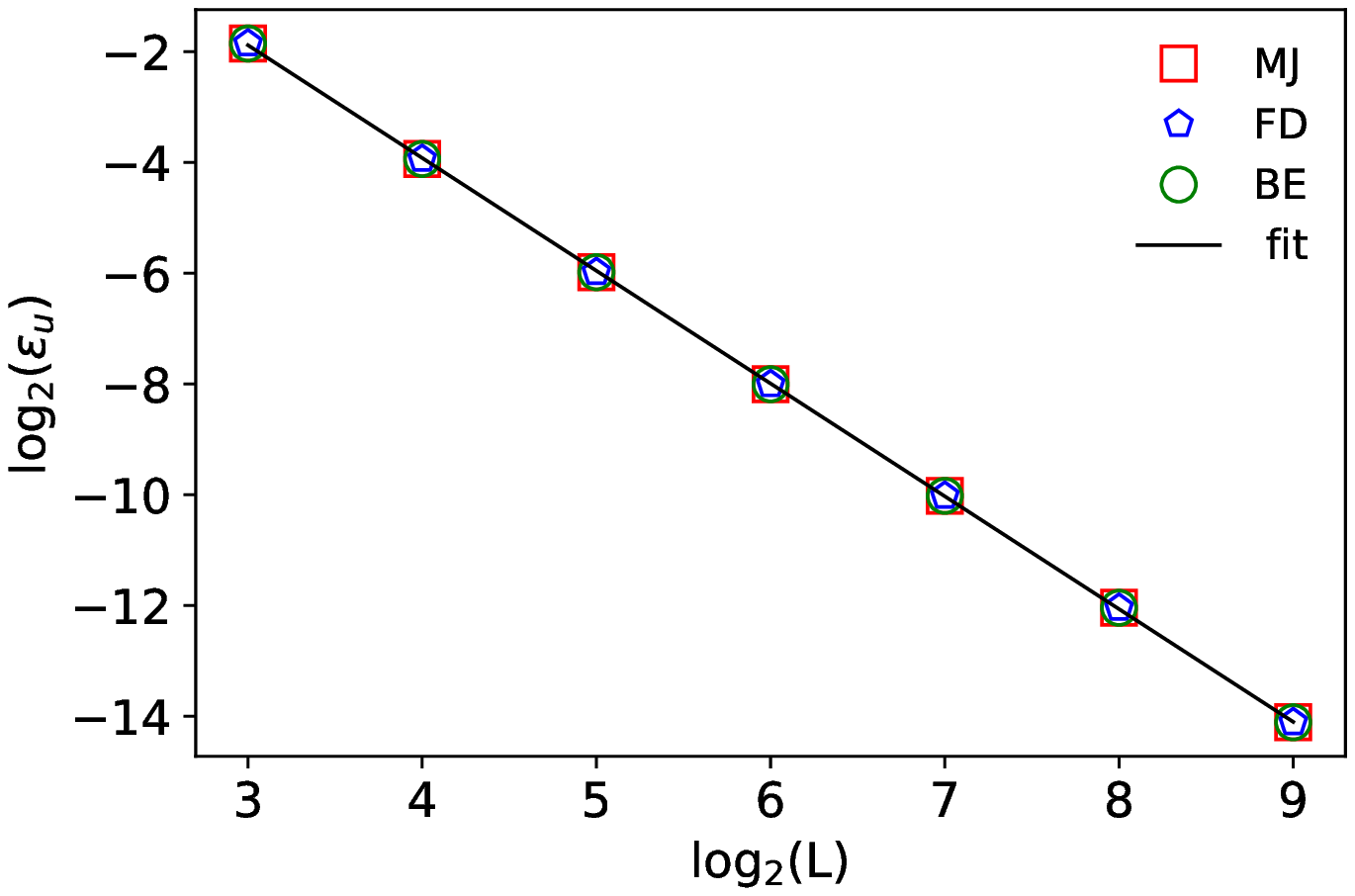}}
\vspace*{8pt}
\caption{Verification of grid convergence in the Poiseuille flow for low velocities. The figure at the top shows the obtained and expected velocity profiles in lattice units for the model based on the MJ distribution for $L=64$. The figure on the bottom show the grid convergence and the linear fit, which gives the same result for the three models: slope = $-2.037$.}
\label{err-fig}
\end{figure}

We verify the grid convergence of our models in the context of Poiseuille flow, which consists in forcing the fluid to flow through two parallel plates. For convenience, we consider small macroscopic velocities, i.e., the fluid is composed by ultra-relativistic particles, but the flow is non-relativistic. Thus, a parabolic $x$-component of the velocity is expected, $u_x^{\text{ref}}(y) = a/(2\nu)(y^2 - yL)$, which is the solution of the macroscopic equations for small velocities and will be used as our reference solution. We use squared domains of different sizes: $L = \{8,\, 16,\, 32,\, 64,\, 128,\, 256,\, 512\}$. The relaxation time is set constant, $\tau = 3$, which correspond to $\nu = 0.6864$ according to Eq.\eqref{nu-ref-eq}. Initially, the fluid is at rest, $\boldsymbol{u}_0=0$, and we have $\rho_0 = 1$ everywhere. A constant external acceleration of magnitude $a = 10^{-8}$ lattice units is applied in the $x$-direction by means of momentum update, $\boldsymbol{u} = \boldsymbol{u} + \boldsymbol{a}\tau$, in analogy to the Shan-Chen forcing scheme in the classical case ($\gamma\approx 1$)~\cite{kruger16}. For a more general relativistic forcing scheme, see Ref.~\cite{romatschke11}. On the top and bottom boundaries, we impose $\boldsymbol{u} =0$, $\rho=1$ and $\theta=1$ in the equilibrium distribution and extrapolate the non-equilibrium distribution from the nearest fluid node (see Ref.~\cite{doi:10.1063/1.1471914}). On the $x$-direction we use periodic boundary conditions. The simulation is stopped when a convergence of $10^{-12}$ is achieved in the velocity field, where the error is calculated by the spatial average of $\vert u_x^{new} - u_x^{old} \vert/\vert u_x^{new} \vert$ at every fluid point where $\vert u_x^{new} \vert >0$. The error of the LBM solution compared to the analytical solution is calculated by:
\begin{flalign*}
&\epsilon _u = \sqrt{\frac{\sum_{y} [ u_x(y) - u_x^{\text{ref}}(y)  ]^2  }{ \sum_{y} [ u_x^{\text{ref}}(y)  ]^2 } }.
\end{flalign*}
As an example, we show in Fig. \ref{err-fig} (top) the solution from the model based on the MJ distribution together with the analytical solution for $L=64$. Notice that the parabolic profile is recovered and that the velocities achieved are small compared to the speed of light ($c=1$), justifying the non-relativistic limit for the analytical solution. On the bottom part of Fig. \ref{err-fig}, one can see that the error decreases exponentially as the system size increases. The same slope is found for the three models by means of a linear fit: $-2.037$. This result demonstrates the grid convergence and shows that our models are second order accurate, as expected for LBMs.

\subsection{Hot spot relaxation}\label{hot-spot-sec}

\begin{figure}[htb]
\centerline{\includegraphics[width=0.9\linewidth]{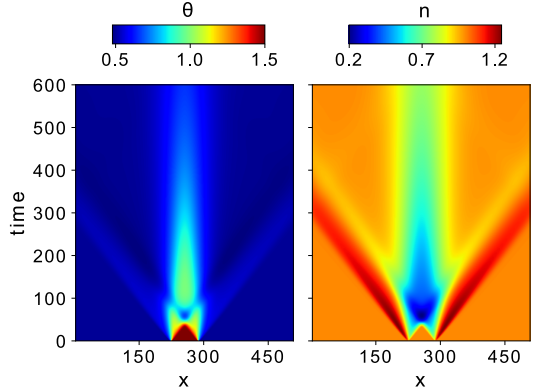}}
\vspace*{8pt}
\caption{Time evolution of the temperature (left), where $\theta=T/T_r$, with $T_r=1.0$, and density (right) at $y=L_Y/2$. The model used is based in a fifth order expansion of MJ distribution.}
\label{hp-fields2-fig}
\end{figure}
\begin{figure}[htb]
\subfloat{%
  \includegraphics[width=.9\linewidth]{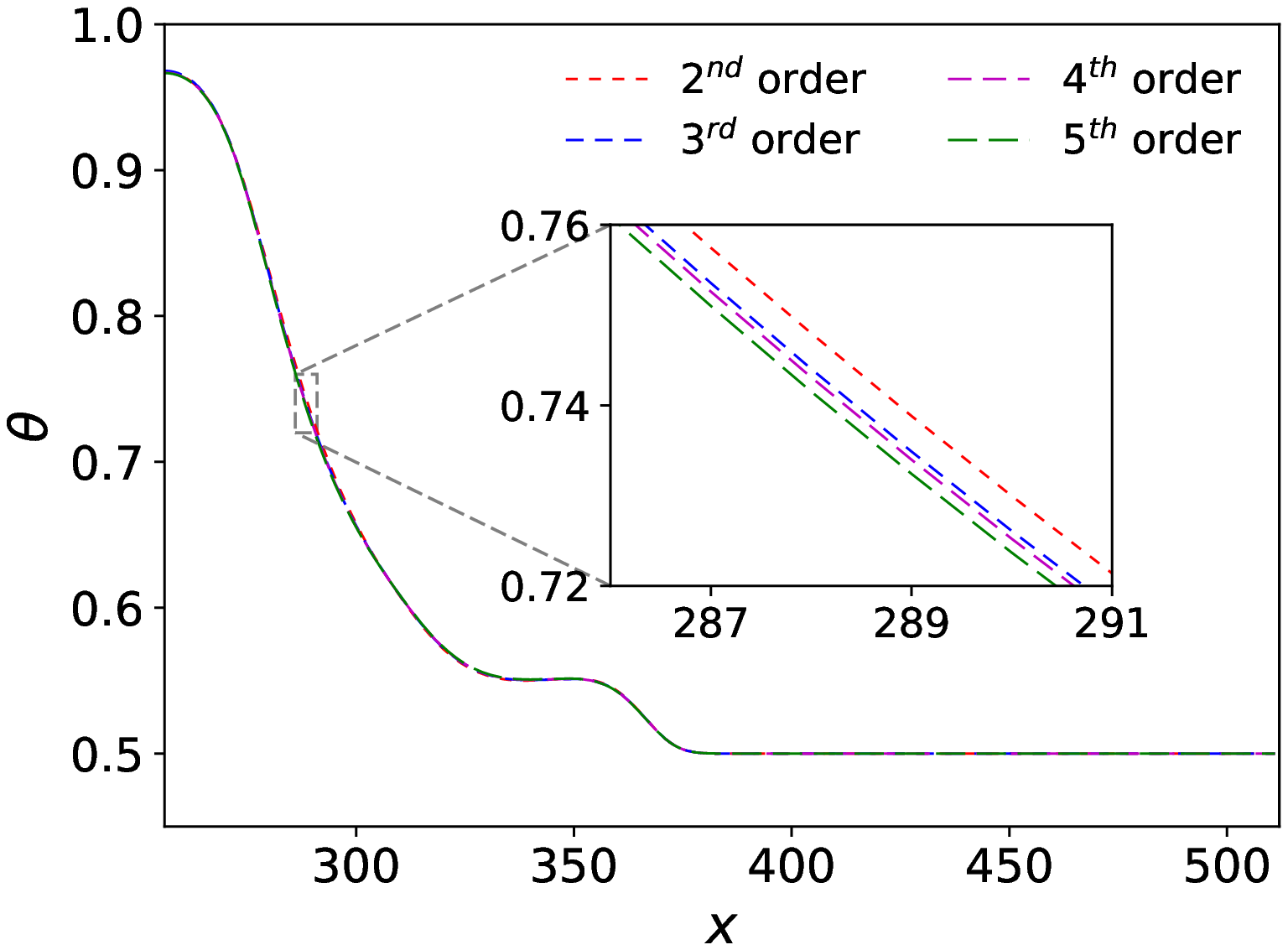}%
}\hfill
\subfloat{%
  \includegraphics[width=.9\linewidth]{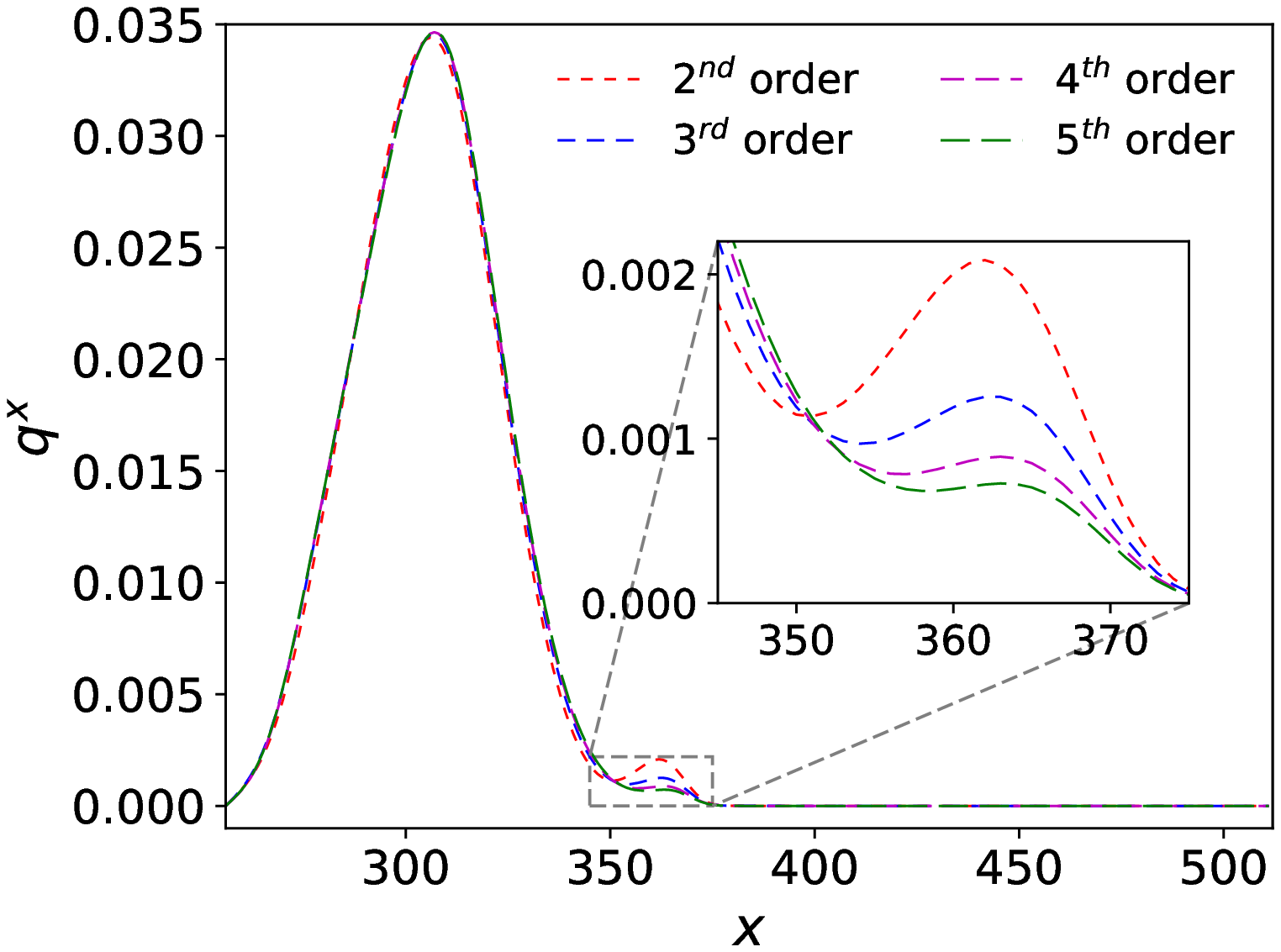}%
}
\caption{Temperature and heat flux profiles at $t=100$ at $y=L_Y/2$ and $256 \leq x < 512 $ for MJ distribution. Due to the symmetry, this region in space contains all relevant information about the problem. The inset shows the differences between different expansion orders.}
\label{hs-t-fig}
\end{figure}

In the hot spot relaxation experiment~\cite{doi:10.1063/1.2919791, PhysRevE.57.978, PhysRevB.92.115426} an homogeneous fluid is heated within a limited region (e.g., with a laser), and then let to relax to equilibrium. Here, we perform this numerical experiment using our model based on the MJ distribution expanded from second to fifth order in the relativistic polynomials. The relaxation time is calculated following the measurements done in Sec. \ref{viscosity-sec}: $\tau = 4 \eta/(s \theta) + 0.2546\delta t $, where $\eta/s = 0.5$ and $\theta$ is the local temperature. In a system with dimensions $L_X \times L_Y = 512 \times 512$, we have initially $n_0=1.0$ and $\mathbf{u_0}=0.0$ everywhere and the temperature is $\theta = 1.5$ inside the region $(x-L_X/2)^2+(y-L_Y/2)^2\leq 32^2$, and $\theta=0.5$ elsewhere. Open boundary conditions are used in both directions. In Fig. \ref{hp-fields2-fig}, we see the time evolution of the temperature and density along the line $y=L_Y/2$ (due to the circular symmetry, this region contains all important information). Fig. \ref{hs-t-fig} shows temperature and the $x$-component of the heat flux (Eq.\eqref{heat-eq2}) profiles at time $t=100 $ and at $y=L_Y/2$. In the insets, we can see the deviations between different expansion orders. In order to quantify the differences, we calculate the average deviation of $\theta$ and $q^x$ with respect to the results obtained with the fifth order model:  
\begin{flalign*}
&\epsilon _\theta^{(N)} = \frac{\sum_{x=x_1}^{x_2} \vert \theta ^{(N)} - \theta^{(5)}  \vert  }{ \sum_{x=x_1}^{x_2}  \theta^{(5)}  }  \\
&\epsilon _{q}^{(N)} = \frac{\sum_{x=x_1}^{x_2} \vert (q^x)^{(N)} - (q^x)^{(5)}  \vert  }{ \sum_{x=x_1}^{x_2}  (q^x)^{(5)}  } ,
\end{flalign*}
where $N$ denotes the order of the model used to calculate the field and $x_1$ and $x_2$ delimit the interval considered to calculate the average. Adopting $x_1=256$ and $x_2=388$, where $x_2$ here was chosen as the limit from where the fields remain constant ($q^x=0$ and $\theta=0.5$), we have the following average deviations: 
\begin{flalign*}
&\epsilon_\theta^{(2)} = 0.224 \% \:\:\:\:\: \epsilon_q^{(2)} = 5.195 \% \\
&\epsilon_\theta^{(3)} = 0.114 \% \:\:\:\:\: \epsilon_q^{(3)} = 1.781 \% \\
&\epsilon_\theta^{(4)} = 0.059 \% \:\:\:\:\: \epsilon_q^{(4)} = 0.744 \%
\end{flalign*}
These deviations are due to the different values that the truncated expansions give for the same set of parameters (temperature, velocity, chemical potential), as can be seem in Fig. \ref{comp-fig} for the equilibrium distribution. This result reinforces that higher expansion order provides a more accurate description, specially in problems with high velocity and/or high temperature deviations.

\section{Summary and Conclusions}\label{conclusion-sec}

We presented new lattice Boltzmann methods for ultra-relativistic fluids in two dimensions governed by the Maxwell-J\"uttner and the Bose-Einstein distributions expanded up to fifth order and studied them together with the model for Fermi-Dirac distribution presented in Ref.~\cite{coelho17}. New polynomials and quadratures were developed using appropriate weight functions, which are the distributions themselves with zero macroscopic velocity. We analyzed the differences between different expansion orders concluding that, as expected, higher order expansions give more accurate results for higher velocities and temperature deviations. Since the fifth order expansion is, so far, the highest used in relativistic LBMs, the models presented here are recommended for simulations where accuracy is important. 

The transport coefficients were numerically measured for our models. The previous models used the viscosity given by Grad's theory, but, as verified with the measurements, this theoretical value is not reproduced by the relativistic fluids. We also measured the kinematic viscosity and thermal conductivity using a reference model, which uses exact streaming and, therefore, gives results without numerical difusivity. The measurement suggests that
\begin{flalign*}
\eta = \frac{(\varepsilon + P)}{4} \left( \tau - \frac{\delta t}{2} \right),\:\:\:\: \mbox{and} \:\:\:\:   \kappa = \frac{3\,\tau \,g_3(z)}{2\, g_2(z)}.
\end{flalign*}
These coefficients should be confirmed in future analytical calculations using Chapman-Enskog expansion. As demonstrated recently~\cite{gabbana17}, relativistic hydrodynamics is better described by Chapman-Enskog rather than by Grad method, which, to the best of our knowledge, has not been done yet for two-dimensional ultra-relativistic fluids. 

The present study opens the way to develop a fully dissipative model with multi-relaxation time collision operator, similarly as done in Ref.~\cite{PhysRevD.86.085044}, in order to independently adjust the transport coefficients and enhance the numerical stability. The presented models can also be extended to three dimensions following similar procedures. The relativistic polynomials can straightforwardly be calculated for three dimensions, since they are written in a tensorial form, and the quadratures can be calculated by finding the roots of the radial and angular polynomials as done here. The EDF expansion becomes significantly more complicated, but one can use the ansatz described in Ref.~\cite{romatschke11}.

\section*{Acknowledgments}

The authors thank to the European Research Council (ERC) Advanced Grant 319968-FlowCCS and to FAPERJ for the financial support.

 \appendix


\section{Integrals used in the expansion}\label{integrals-sec}

Here we show more details about the calculation of the integrals used in the EDF expansion, Eq.\eqref{expansion-eq}, which are laborious to solve analytically even using computer algorithms. Consider, for instance, the projection in the second order polynomial $P^{i_1i_2}$,
\begin{flalign}\label{proj-2-ex-eq}
&A^{i_1i_2} \\
&= \int  \frac{d^2 p}{p^0} \frac{D_1p^{i_1}p^{i_2}+[D_2(p^0)^2+D_3p^0+D_4]\delta^{i_1i_2}}{z^{-1}e^{p^0\gamma(1-\mathbf{v}\cdot\mathbf{u})/\theta)}+\xi}.\nonumber
\end{flalign}
Because of the inner product $\mathbf{v}\cdot\mathbf{u}$ (where $\mathbf{v}=\mathbf{p}/|\mathbf{p}|$) in the exponential, the integration in separate spatial components is more complicated than in non-relativistic case. To solve this kind of integral we consider the integration of each monomial separately, which, in the example above, are $\{p^xp^x,\, p^xp^y,\, p^yp^y,\,(p^0)^2, \, p^0,\, 1\}$.   

One can write $\mathbf{v}\cdot\mathbf{u} = u \cos(\phi-\alpha)$ where $\alpha$ is the angle between $\mathbf{u}$ and the x-axis and $\phi$ is the angle between $\mathbf{v}$ and the x-axis. Lets define a generic integral of $f^{eq}$ as
\begin{flalign*}
I_1^{(mnq)}\equiv \int^{2\pi}_0 \int^{\infty}_0 dp d\beta \frac{p^m\sin^n(\beta)\cos^q(\beta)}{z^{-1}e^{p\gamma(1-u\cos(\beta)/\theta)}+\xi},
\end{flalign*}
where $\beta\equiv \phi-\alpha$. If $n$ is odd, the integral is zero, but if $n$ is even
\begin{flalign*}
I_1^{(mnq)}&= 2\int^{\pi}_0 \int^{\infty}_0 dp d\beta \frac{p^m\sin^n(\beta)\cos^q(\beta)}{z^{-1}e^{p\gamma(1-u\cos(\beta)/\theta)}+\xi}.
\end{flalign*}
Using the identity  
\begin{flalign*}
\int^\infty_0 dp\frac{p^m}{z^{-1}e^{p \,y}+\xi} = y^{-(m+1)}\Gamma(m+1) g_{m+1}(z) ,
\end{flalign*}
where $g_\nu(z)$ is defined in Eq.\eqref{gz-eq}, and after some algebra, we find
\begin{flalign}\label{I1-eq}
I_1^{(mnq)} &= \frac{2\theta^{m+1}}{\gamma^{m+1}u^{n+q}}\Gamma(m+1)g_{m+1}(z)\\ \nonumber
&\cdot\int^{\pi}_0 dw [u^2-(1-w)^2]^{(n-1)/2}\frac{(1-w)^q}{w^{m+1}}.
\end{flalign}
The remaining integral can be solved exactly for given integers $m$, $n$ and $q$. With this, we have separated the original integral of two variables in two non-dimensional integrals in one variable, including the EDF integral $g_\nu(z)$. 

The Eq.\eqref{I1-eq} allows us to calculate projections of the EDFs in the monomials:
\begin{flalign*}
I_2^{mnq}\equiv \int\frac{d^2 p}{p^0} \frac{p^m (p^x)^n (p^y)^q}{z^{-1}e^{p\gamma(1-\mathbf{v}\cdot\mathbf{u})/\theta)}+\xi}.
\end{flalign*}
As an example, lets consider the projection in the monomial $p^xp^x$,
\begin{flalign*}
I_2^{020}&=\int  \frac{d^2 p}{p^0} \frac{(p^x)^2}{z^{-1}e^{p\gamma(1-\mathbf{v}\cdot\mathbf{u})/\theta)}+\xi}\\ \nonumber
&= \int^{2\pi}_0 \int^{\infty}_0 dp d\phi \frac{(p \cos(\beta+\alpha))^2}{z^{-1}e^{p\gamma(1-u\cos(\beta))/\theta)}+\xi}
\end{flalign*}
Since 
\begin{flalign*}
&\cos^2(\alpha+\beta)=\frac{1}{2}+\frac{1}{2}\cos^2(\alpha)\cos^2(\beta)\\ \nonumber &-\frac{1}{2}\cos^2(\beta)\sin^2(\alpha) -2\cos(\alpha)\sin(\alpha)\cos(\beta)\sin(\beta)\\ \nonumber &-\frac{1}{2}\cos^2(\alpha)\sin^2(\beta)+\frac{1}{2}\sin^2(\alpha)\sin^2(\beta),
\end{flalign*}
and considering that the odd powers of $\sin(\beta)$ give null integrals, we write
\begin{flalign*}
I_2^{020} = \frac{1}{2}I_1^{(200)} + \frac{1}{2}(\cos^2(\alpha)-\sin^2(\alpha))(2I_1^{(202)}-I_1^{(200)}).
\end{flalign*}
Thus, using Eq.\eqref{I1-eq}, we have
\begin{flalign*}
I_2^{020} = 2\pi \theta^3 g_3(z)\gamma^2 [(1-u^2)+3 u^x u^x].
\end{flalign*}
And with the integrals $I_2$, one can calculate any projection of the EDF. Eq.\eqref{proj-2-ex-eq}, for instance, becomes for the $A^{xx}$ component:
\begin{flalign*}
A^{xx} = D_1 I_2^{020} + D_2 I_2^{200} + D_3 I_2^{100} + D_4 I_2^{000}.
\end{flalign*}
Below we see all the integrals needed to perform the fifth order expansion
\begin{flalign*}
I_2^{000 } &= 2 \pi \theta g_1(z)\\
I_2^{100 } &= 2 \pi \theta^2 g_2(z) \gamma
\\
I_2^{200 } &= 2 \pi \theta^3 g_3(z) (2 + u^2) \gamma^2
\\
I_2^{300 } &= 6 \pi \theta^4 g_4(z) (2 + 3 u^2) \gamma^3
\\
I_2^{400 } &= 6 \pi (8 + 24 u^2 + 3 u^4) \theta^5 g_5(z) \gamma^4
\\
I_2^{010 } &= 2 \pi \theta^2 g_2(z) u^x \gamma
\\
I_2^{020 } &= 2 \pi \theta^3 g_3(z) (1 - u^2 + 3 u^x u^x) \gamma^2
\\
I_2^{030 } &= 6 \pi \theta^4 g_4(z) (3 (1 - u^2) u^x + 5 (u^x)^3) \gamma^3
\\
I_2^{040 } &= (6 \pi (3 + 3 u^4 + 30 (u^x)^2 + 35 (u^x)^4 - 
       6 u^2 (1 \\ &+ 5 (u^x)^2)) \theta^5 g_5(z)) \gamma^4
       \\
I_2^{001 } &= 2 \pi \theta^2 g_2(z) u^y \gamma
\\
I_2^{002 } &= 2 \pi \theta^3 g_3(z) (1 - u^2 + 3 u^y u^y) \gamma^2
\\
I_2^{003 } &= 6 \pi \theta^4 g_4(z) (3 (1 - u^2) u^y + 5 (u^y)^3) \gamma^3
\\
I_2^{004 } &= ((6 \pi (3 + 3 u^4 + 30 (u^y)^2 + 35 (u^y)^4 - 
        6 u^2 (1 \\ & + 5 (u^y)^2)) \theta^5 g_5(z)) \gamma^4)
        \\
I_2^{220 } &= 6 \pi (4 - 3 u^2 - u^4 + 
     5 (u^x)^2 (6 + u^2)) \theta^5 g_5(z) \gamma^4
     \\
I_2^{120 } &= 6 \pi \theta^4 g_4(z) ((1 - u^2) + 5 u^x u^x) \gamma^3
\\
I_2^{202 } &= 6 \pi (4 - 3 u^2 - u^4 + 
     5 (u^y)^2 (6 + u^2)) \theta^5 g_5(z) \gamma^4
     \\
I_2^{102 } &= 6 \pi \theta^4 g_4(z) ((1 - u^2) + 5 u^y u^y) \gamma^3
\\
I_2^{022 } &= (6 \pi ((1 - u^2) + 4 (1 - u^2) u^2 \\ &+ 
       35 (u^x)^2 (u^y)^2) \theta^5 g_5(z)) \gamma^4
       \\
I_2^{031 } &= 30 \pi \theta^5 g_5(z) (3 (1 - u^2) + 7 (u^x)^2) u^x u^y \gamma^4
\\
I_2^{211 } &= 30 \pi (6 + u^2) \theta^5 g_5(z) u^x u^y \gamma^4
\\
I_2^{011 } &= 6 \pi \theta^3 g_3(z) u^x u^y \gamma^2
\\
I_2^{111 } &= 6 \pi \theta^4 g_4(z) (5 u^x u^y) \gamma^3
\\
I_2^{130 } &= -30 \pi u^x (-3 + 3 u^2 - 7 (u^x)^2) \theta^5 g_5(z) \gamma^4
\\
I_2^{310 } &= 30 \pi (4 + 3 u^2) \theta^5 u^x g_5(z) \gamma^4
\\
I_2^{110 } &= 6 \pi \theta^3 g_3(z) u^x \gamma^2
\\
I_2^{101 } &= 6 \pi \theta^3 g_3(z) u^y \gamma^2
\\
I_2^{013 } &= 30 \pi \theta^5 g_5(z) (3 (1 - u^2) + 7 (u^y)^2) u^x u^y \gamma^4\\
I_2^{103 } &= -30 \pi u^y (-3 + 3 u^2 - 7 (u^y)^2) \theta^5 g_5(z) \gamma^4
\\
I_2^{121 } &= -30 \pi (-1 + u^2 - 7 (u^x)^2) u^y \theta^5 g_5(z) \gamma^4
\\
I_2^{201 } &= 6 \pi \theta^4 g_4(z) (4 + u^2) u^y \gamma^3
\\
I_2^{301 } &= 30 \pi (4 + 3 u^2) \theta^5 u^y g_5(z) \gamma^4
\\
I_2^{021 } &= 6 \pi \theta^4 g_4(z) ((1 - u^2) + 5 u^x u^x) u^y \gamma^3
\\
I_2^{112 } &= -30 \pi (-1 + u^2 - 7 (u^y)^2) u^x \theta^5 g_5(z) \gamma^4
\end{flalign*}
\begin{flalign*}   
I_2^{210 } &= 6 \pi \theta^4 g_4(z) (4 + u^2) u^x \gamma^3
\\
I_2^{012 } &= 6 \pi \theta^4 g_4(z) ((1 - u^2) + 5 u^y u^y) u^x \gamma^3
\\
I_2^{050 }&= 30 \pi u^x (15 (1 - u^2)^2 + 70 (u^x)^2 (1 - u^2) \\ &+ 
     63 (u^x)^4) \theta^6 g_6(z) \gamma^5\\
I_2^{005 }&= 30 \pi u^y (15 (1 - u^2)^2 + 70 (u^y)^2 (1 - u^2)\\ & + 
     63 (u^y)^4) \theta^6 g_6(z) \gamma^5\\
I_2^{500 }&= 30 \pi (8 + 40 u^2 + 15 u^4) \theta^6 g_6(z) \gamma^5\\
I_2^{230 }&= 30 \pi u^x (18 - 3 u^4 + 56 (u^x)^2 \\ &+ 
     u^2 (-15 + 7 (u^x)^2)) \theta^6 g_6(z) \gamma^5\\
I_2^{203 }&= 30 \pi u^y (18 - 3 u^4 + 56 (u^y)^2 \\ &+ 
     u^2 (-15 + 7 (u^y)^2)) \theta^6 g_6(z) \gamma^5\\
I_2^{041 }&= 90 \pi ((1 - u^2)^2 + 14 (u^x)^2 (1 - u^2) \\ &+ 
     21 (u^x)^4) u^y \theta^6 g_6(z) \gamma^5\\
I_2^{221 }&= -30 \pi (-6 (1 - u^2) - u^2 (1 - u^2) - 56 (u^x)^2 \\ &+ 
     u^2 (-7 (u^x)^2)) u^y \theta^6 g_6(z) \gamma^5\\
I_2^{014 }&= 90 \pi ((1 - u^2)^2 + 14 (u^y)^2 (1 - u^2)\\ & + 
     21 (u^y)^4) u^x \theta^6 g_6(z) \gamma^5\\
I_2^{032 }&= -30 \pi u^x ((u^x)^2 (-1 + 4 u^2 - 63 (u^y)^2) \\ &+ 
     3 (-1 + (-5 + 6 u^2) (u^y)^2)) \theta^6 g_6(z) \gamma^5\\
I_2^{212 }&= -30 \pi (-6 (1 - u^2) - u^2 (1 - u^2) - 56 (u^y)^2 \\ &+ 
     u^2 (-7 (u^y)^2)) u^x \theta^6 g_6(z) \gamma^5\\
I_2^{410 }&= 90 \pi u^x (8 + 12 u^2 + u^4) \theta^6 g_6(z) \gamma^5\\
I_2^{023 }&= -30 \pi u^y ((u^y)^2 (-1 + 4 u^2 - 63 (u^x)^2) \\ &+ 
     3 (-1 + (-5 + 6 u^2) (u^x)^2)) \theta^6 g_6(z) \gamma^5\\
I_2^{401 }&= 90 \pi u^y (8 + 12 u^2 + u^4) \theta^6 g_6(z) \gamma^5
\\
I_2^{140 }&= 90 \pi ((1 - u^2)^2 + 14 (u^x)^2 (1 - u^2) \\ &+ 
     21 (u^x)^4) \theta^6 g_6(z) \gamma^5\\
I_2^{320 }&= -30 \pi (-4 (1 - u^2) - 3 u^2 (1 - u^2) - 42 (u^x)^2 \\ &+ 
     u^2 (-21 (u^x)^2)) \theta^6 g_6(z) \gamma^5\\
I_2^{104 }&= 90 \pi ((1 - u^2)^2 + 14 (u^y)^2 (1 - u^2) \\ &+ 
    21 (u^y)^4) \theta^6 g_6(z) \gamma^5\\
I_2^{302 }&= -30 \pi (-4 (1 - u^2) - 
    3 u^2 (1 - u^2) \\ & - 42 (u^y)^2 + u^2 (-21 (u^y)^2)) \theta^6 g_6(z) \gamma^5\\
I_2^{122 }&= -30 \pi (-1 + u^2 - 6 u^2 (1 - u^2)\\& - 
     63 (u^x)^2 (u^y)^2) \theta^6 g_6(z) \gamma^5\\
I_2^{131 }&= -630 \pi u^x (-1 + u^2 - 3 (u^x)^2) u^y \theta^6 g_6(z) \gamma^5\\
I_2^{311 }&= 315 \pi (2 + u^2) \theta^6 g_6(z) 2 u^x u^y \gamma^5\\
I_2^{113 }&= -630 \pi u^y (-1 + u^2 - 3 (u^y)^2) u^x \theta^6 g_6(z) \gamma^5
\end{flalign*}

\bibliographystyle{elsarticle-num} 
\bibliography{references}


\end{document}